\documentclass[a4paper,11pt]{article}
\usepackage{jinstpub} 
\usepackage{lineno}
\usepackage{rotating}
\usepackage{booktabs}
\newcommand\pubnumber{SLAC-PUB-17734}
\newcommand\pubdate{\today}
\newcommand\pubblock{\rightline{\begin{tabular}{l} \pubnumber \\
\pubdate \end{tabular}}}

\title{\boldmath XCC: An X-ray FEL-based $\gamma\gamma$ Compton Collider Higgs Factory}

\author{T. Barklow,}
\author{C. Emma,}
\author{Z. Huang,}
\author{A. Naji,}
\author{E. Nanni,} 
\author{A. Schwartzman,}
\author{S. Tantawi,}
\author{and G. White}

\affiliation{SLAC National Accelerator Laboratory, Menlo Park, California, U.S.A.}

\emailAdd{timb@slac.stanford.edu}

\abstract{  This report describes the conceptual design of a $\gamma\gamma$ Higgs factory in which 
  $62.8$~GeV electron beams collide with  1~keV X-ray free electron laser (XFEL) beams to produce colliding beams of 62.5~GeV photons. The Higgs boson production rate is 
  80,000 Higgs bosons per $10^7$ second year, roughly the same as the ILC Higgs rate at $\sqrt{s}=250$~GeV. The electron accelerator is based on cold copper distributed coupling (C$^3$) accelerator  technology. Unlike the center-of-mass energy spectra of previous optical wavelength $\gamma\gamma$ collider designs, the sharply peaked $\gamma\gamma$ center-of-mass energy spectrum of XCC produces model independent Higgs coupling measurements with precision on par with $e^+e^-$ colliders.  For the triple Higgs coupling measurement, the XCC 
center-of-mass energy can be upgraded to 380 GeV, where the cross section for $\gamma\gamma\rightarrow HH$ is twice that of  $e^+e^- \rightarrow ZHH$ at $\sqrt{s}=500$~GeV.   Design challenges are discussed, along with the R\&D to address them, including demonstrators.}


\begin{document}

\begin{flushright}
\pubblock
\end{flushright}

\maketitle
\flushbottom

\section{Introduction}

The staging of an electron-positron linear collider with an initial $\gamma\gamma$ collider at the Higgs resonance is not a new idea.  For example, former KEK director Hirotaka Sugawara proposed such a staging sequence  for the Internatioal Linear Collider (ILC) in 2009.  The idea was ultimately rejected due in part to a weak physics case. However, with an X-ray laser in place of an optical laser, the physics case for a first stage $\gamma\gamma$ collider is strengthened considerably. In fact,  the optimum second stage for such a facility could again be a $\gamma\gamma$ collider, this time operating 
at $\sqrt{s}=380$~GeV to produce $\gamma\gamma\rightarrow HH$ for the study of the Higgs self-coupling.  Due to its lower cost and smaller footprint, the XCC could begin operation on an earlier timescale than an $e^+e^-$ Higgs factory.  Indeed, the lower cost could be  critical in an era  with many other commitments by government funding agencies


\subsection{Concept Overview}
To date, $\gamma\gamma$ collider Higgs factory designs have utilized optical wavelength lasers\cite{Ginzburg:1981vm}\cite{Ginzburg:1982yr}\cite{Telnov:1989sd}\cite{Asner:2001ia}\cite{Asner:2001vh}.  The
center-of-mass energy of the electron--photon system is usually constrained to $x<4.82$, where $x=4E_{e}\omega_0/m^2_e$, $m_e$ is the electron mass and $E_{e}$ ($\omega_0$) is the electron (laser photon) energy.  Larger $x$ values are problematic due to the linear QED thresholds of $x=4.82$  ($x=8.0$) for the processes $\gamma\gamma_0\rightarrow e^+e^-$  ($e^-\gamma_0\rightarrow e^-e^+e^-$), where $\gamma$ and $\gamma_0$ refer to the Compton-scattered and laser photon, respectively.  Larger $x$ values, however, also carry advantages.  As $x$ is increased,
the $\gamma\gamma$  luminosity distribution with respect to center-of-mass energy is more sharply peaked near the maximum center-of-mass energy value. Such a distribution increases the production rate of a narrow resonance relative to $\gamma\gamma$ background processes when the peak 
is tuned to the resonance mass.

\begin{figure}
\centering
     \includegraphics[width=1.00\textwidth]{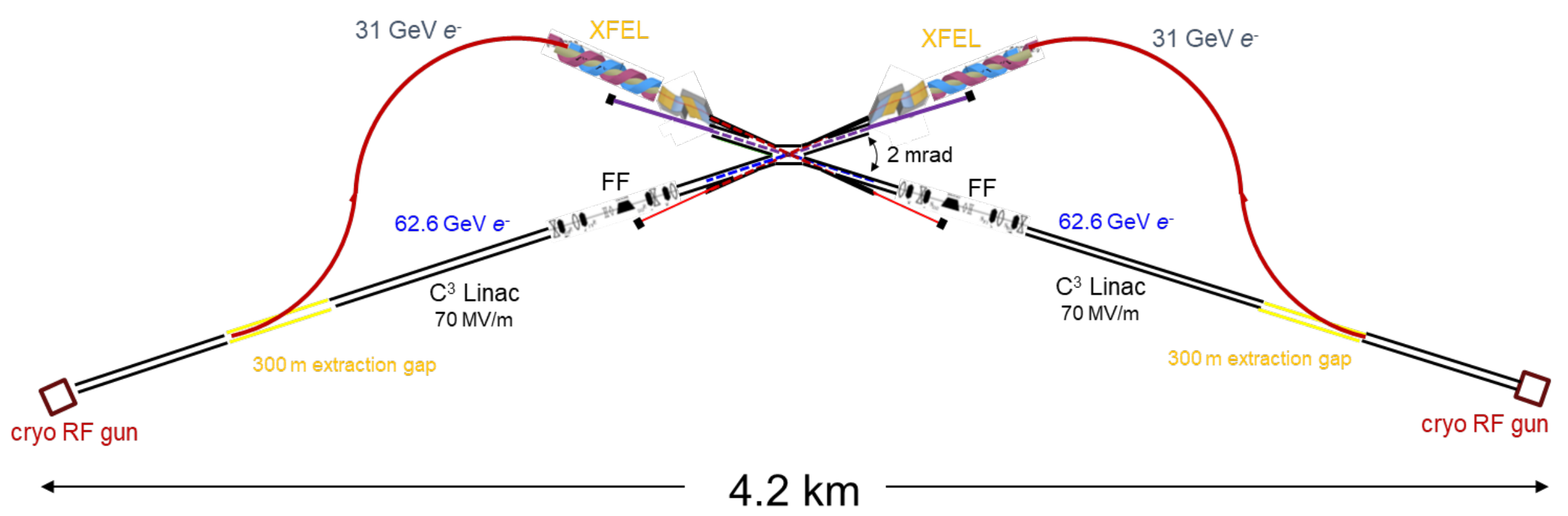}
     \caption{Schematic of XCC including cryogenic RF injector, C$^3$ Linac, electron beam final focus (FF) and XFEL.}
     \label{fig:schematic}
 \end{figure}

A schematic of the $\gamma\gamma$ collider, or XFEL Compton Collider (XCC),
is shown in figure~\ref{fig:schematic}.   A low emittance cryogenic RF gun produces 90\% polarized electrons with $0.62\times 10^{10}$ electrons per bunch and 330 bunches per train at a repetition
rate of 120~Hz.   The normalized horizontal and vertical gun emittances are 0.12 microns each.  A linear accelerator (Linac) utilizing  cold copper distributed coupling (C$^3$) technology\cite{Bane:2018fzj}\cite{Dasu:2022nux} accelerates the electron bunches with an interaction point bunch spacing of 4.2 ns and a gradient of 70 MeV/m.
At the 31~GeV point, every other bunch is diverted to the XFEL line where a helical undulator produces circularly polarized 1~keV X-ray light with 0.7 Joules per pulse. In order to limit coherent synchrotron radiation, a 300 meter extraction gap is added to each accelerator at the 31~GeV point, which increases the extraction line radius of curvature to 133~km.  The remaining bunches continue down the Linac until reaching an energy of $E_{e^-}=62.8$~GeV, after which they pass through a final focus section that squeezes the geometric horizontal and vertical spot sizes to 5.4~nm at the primary
$e^-e^-$ interaction point (IP).  The $e^-e^-$ geometric luminosity is $2.1\times 10^{35} \rm{cm}^2\ \rm{s}^{-1}$.  
At the Compton interaction point (IPC), the 62.8 GeV electrons collide with the X-ray laser light from the opposing XFEL line to produce 62.5~GeV photons which collide at the IP to produce Higgs bosons with mass $m_H=125$~GeV.  The X-ray light has been focused at this point from $9\ \mu$m at the end of the XFEL to a waist radius of $a_\gamma$=30~nm using boron carbide coated Kirkpatrick-Baez (KB) mirrors.  The total length of 4.2~km is chosen to accommodate an energy upgrade to $\sqrt{s}=380$~GeV with a 120~MV/m accelerating gradient.

\begin{figure}
\centering
     \includegraphics[width=0.75\textwidth]{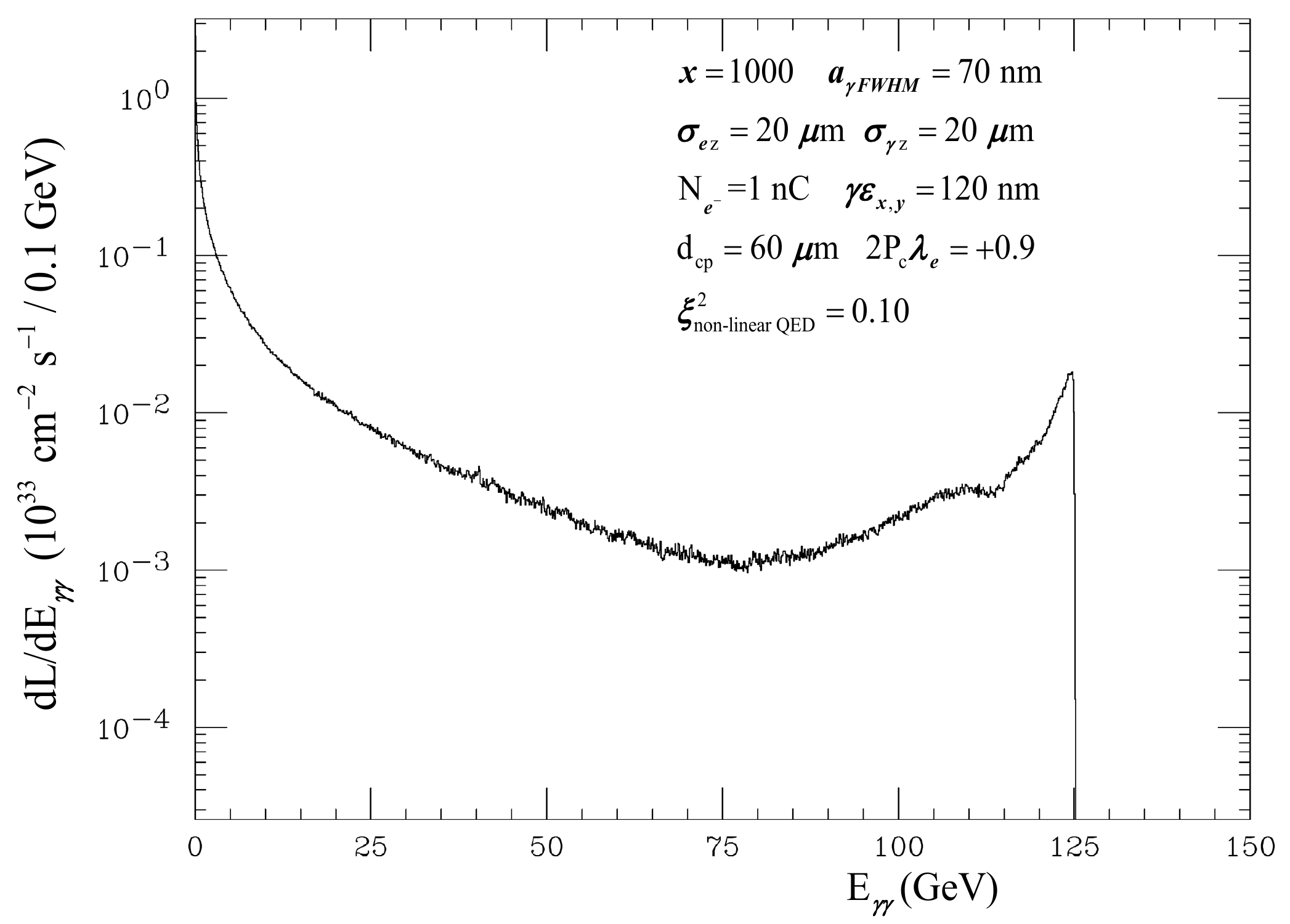}
     \includegraphics[width=0.75\textwidth]{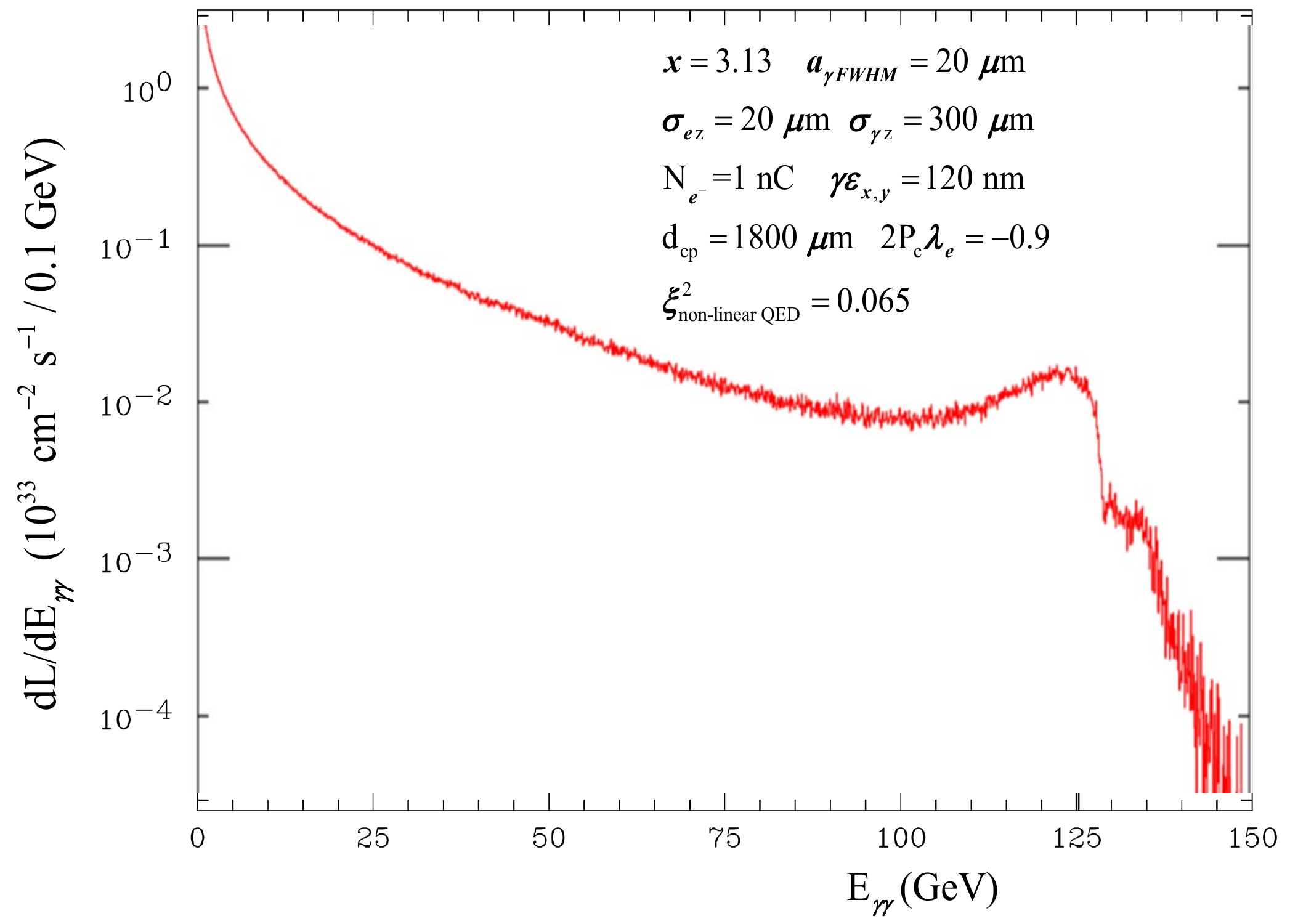}
     \caption{$\gamma\gamma$ luminosity as a function of $\gamma\gamma$ center-of-mass energy $E_{\gamma\gamma}$ for x=1000 \& $2P_c\lambda_e=+0.9$ (top) versus x=3.13 \& $2P_c\lambda_e=-0.9$ (bottom) as calculated by the  CAIN MC, where $P_c$ and $\lambda_e$ are the helicities of the laser photon and electron, respectively. The x-ray waist radius $a_\gamma$ at the Compton interaction point and the electron (x-ray) beam r.m.s longitudinal sizes, $\sigma_{ez}$ ($\sigma_{\gamma z}$), are indicated.  The 0.10\% electron beam energy spread, linear QED Bethe-Heitler scattering  ($e^-\gamma_0\rightarrow e^-e^+e^-$) and non-linear QED effects in Compton scattering ($e^-\gamma_0\rightarrow e^-\gamma$) and Breit-Wheeler scattering ($\gamma\gamma_0\rightarrow e^+e^-$) are included in the CAIN simulation, where $\gamma$ and $\gamma_0$ refer to the Compton-scattered and laser photon, respectively.
     }
     \label{fig:x1000lumiplus}
 \end{figure}

The distribution of $\gamma\gamma$ luminosity versus $\gamma\gamma$ center-of-mass energy $E_{\gamma\gamma}$ as calculated with the CAIN Monte Carlo\cite{Chen:1994jt} is shown in figure~\ref{fig:x1000lumiplus}
for $2P_c\lambda_e=+0.9$, where $P_c=+1$ and $\lambda_e=+0.45$ are the helicities of the  laser photon and electron, respectively. For comparison, the corresponding distribution from an x=3.13 optical laser $\gamma\gamma$ collider (OCC) is also shown.  The OCC --  presented here solely as an optical laser $\gamma\gamma$ collider counter-example to the XCC -- has the same parameters as XCC except that the XFEL is replaced with the optical laser in~\cite{Telnov_2020}, the electron beam energy is increased from 62.8~GeV to 86.5~GeV,  the distance $d_{cp}$ between the IPC and IP has been increased from $60\ \mu$m to $1800\ \mu$m, and $2P_c\lambda_e=-0.9$.  The  distribution for x=1000 has an asymmetric peak at the Higgs boson mass with a leading edge width of 0.3~GeV.  In contrast, the $x=3.13$ distribution has a peak at the Higgs boson mass with a leading edge width of 3.5~GeV and a long high-side tail due to multi-photon non-linear QED Compton scattering, characterized by the parameter $\xi^2=2n_\gamma r_e^2\lambda/\alpha$  where $n_\gamma$ is the laser photon density, $r_e$ is the classical electron radius, and $\lambda$ is the laser photon wavelength.  Although the non-linear QED parameter $\xi^2=0.10$ for the $x=1000$ configuration is 50\% larger than that of the $x=3.13$ configuration, the long high-side non-linear QED tail is absent because the 
difference between the maximum linear and non-linear QED photon energies, $E_e/(x+1)$, is very small.   The large low-side tail in the luminosity distributions for $0<E_{\gamma\gamma}<100$~GeV is due to beamstrahlung radiation.

\subsection{Experimental Environment}

\begin{figure}
\centering
     \includegraphics[width=0.75\textwidth]{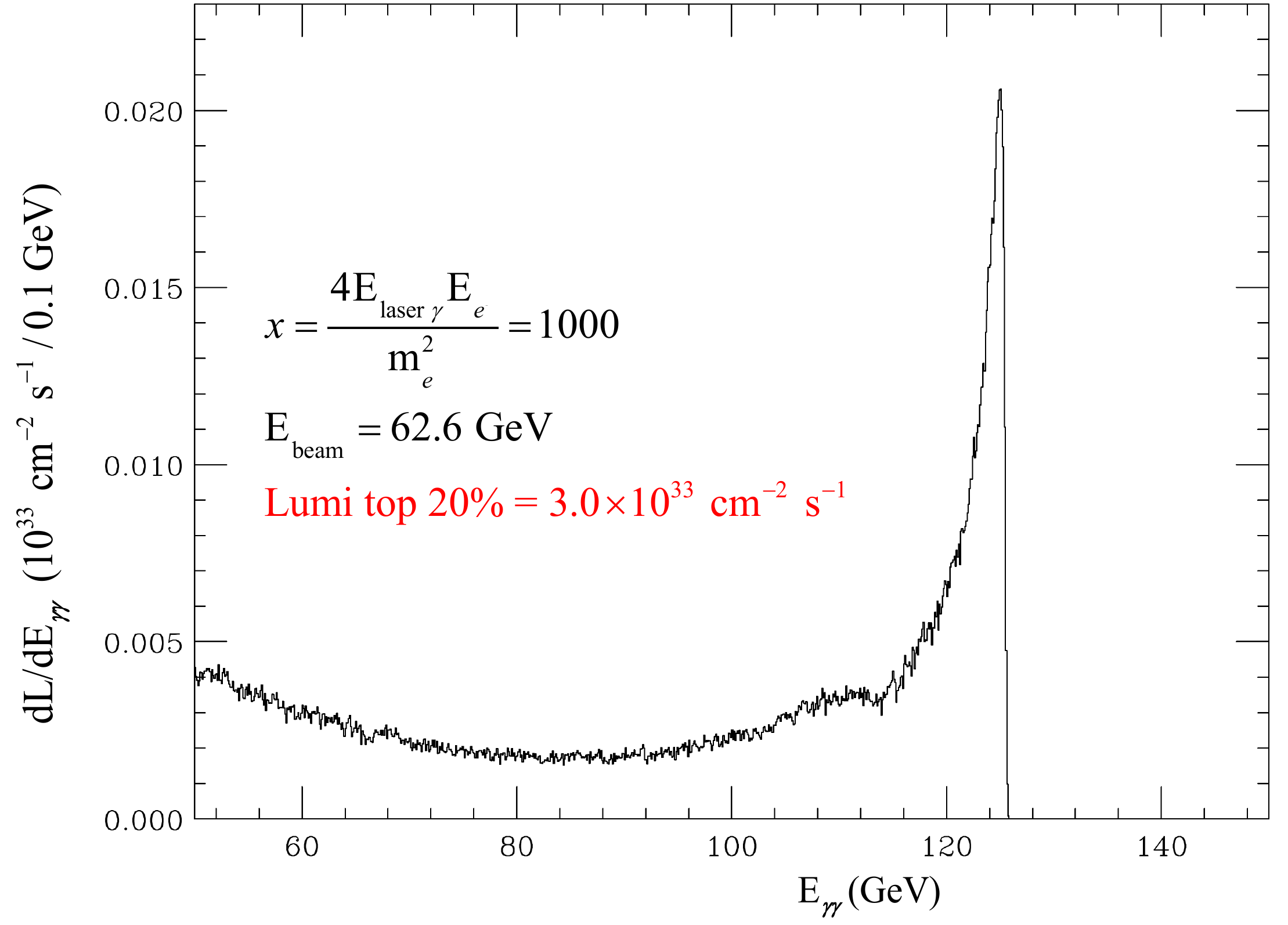}
     \includegraphics[width=0.75\textwidth]{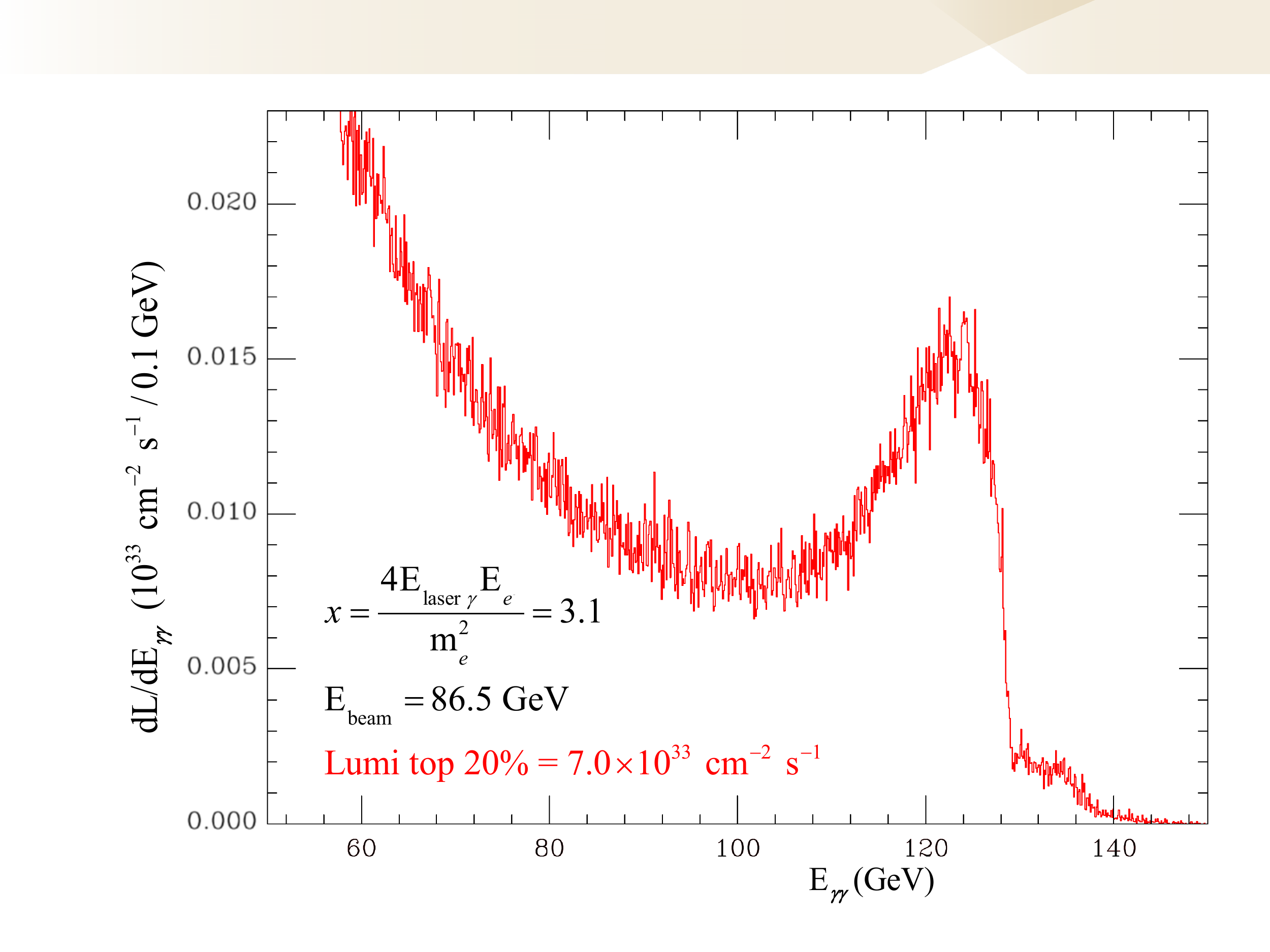}
     \caption{$\gamma\gamma$ luminosity for $\gamma\gamma$ center-of-mass energies $E_{\gamma\gamma}>50$~GeV for x=1000 \& $2P_c\lambda_e=+0.9$ (top) versus x=3.13 \& $2P_c\lambda_e=-0.9$ (bottom) as calculated by the  CAIN MC, where $P_c$ and $\lambda_e$ are the helicities of the laser photon and electron, respectively.  The 0.1\% electron beam energy spread, linear QED Bethe-Heitler scattering  ($e^-\gamma_0\rightarrow e^-e^+e^-$) and non-linear QED effects in Compton scattering ($e^-\gamma_0\rightarrow e^-\gamma$) and Breit-Wheeler scattering ($\gamma\gamma_0\rightarrow e^+e^-$) are included in the CAIN simulation, where $\gamma$ and $\gamma_0$ refer to the Compton-scattered and laser photon, respectively.
     }
     \label{fig:x1000linearlumiplus}
 \end{figure}

The luminosity in figure~\ref{fig:x1000lumiplus} includes a large amount of beamstrahlung
luminosity at low values of $E_{\gamma\gamma}$.  The low transverse momentum $\gamma\gamma\rightarrow\textrm{ hadrons}$ pileup events
produced with this luminosity have low visible energy and are highly boosted in the
forward direction.  For physics analyses, the more relevant
luminosity spectrum is shown in the upper plot of figure~\ref{fig:x1000linearlumiplus}.  With  the XCC, the relatively low luminosity in the top 20\% of  $\gamma\gamma$ center-of-mass energies  is concentrated in one spike where Higgs boson production (proportional to the height of the spike and $m_H/2E_{e^-}$) can be maximized with respect to background. In contrast, the optical laser OCC $\gamma\gamma$ collider (lower plot of figure~\ref{fig:x1000linearlumiplus})  has a larger electron beam energy (87~GeV vs. 63~GeV), a larger top 20\% luminosity, and a smaller peak luminosity at the Higgs resonance.

 Since only 20\% of the electrons at the XCC are converted to photons in the Compton collision, there is also substantial luminosity from $e^-\gamma$,  $\gamma e^-$,  and $e^-e^-$ collisions.
The luminosity for these processes is  listed in table~\ref{tab:lumisummary}.
 
 A comparison of the background rate for ILC, XCC and OCC is shown in table~\ref{tab:higgsratesummary}, where background is defined as the number of hadron events with $\sqrt{\hat{s}/m_H^2}>0.64$. For XCC and OCC, the QCD radiative corrections to the total cross section for $\gamma\gamma\rightarrow q\bar{q}(g)$ have been calculated using equation~4.5 and table~I of reference\cite{Jikia:1996bi}, where  logarithmic extrapolations of the QCD corrections in table~I are used for $s/4m^2 > 2500$.  The XCC background rate is considerably smaller than that of the optical $\gamma\gamma$ collider OCC, and comparable to the ILC background rate.
 
 
 XCC events with $\sqrt{\hat{s}}<100$~GeV are produced with beamstrahlung luminosity and 
 have a distinctive, boosted topology along the beam axis. If such events 
 turn out to be problematic, their rate can be greatly reduced at the cost of some loss in Higgs rate by switching to  asymmetric electron beams.

 \begin{table}

    \centering
    \begin{tabular}{    c  | c | c  } \toprule
        \multicolumn{3}{ c}{$\gamma\gamma$ mode $\sqrt{s}=125$~GeV} \\ \midrule
       & \multicolumn{2}{ c}{Luminosity ($10^{34}$ cm$^{-2}$ s$^{-1}$)} \\[3pt]
        Process  & Total &  $\sqrt{\hat{s}}>100$~GeV \\ \midrule
     $\gamma\gamma$  & 4.7 & 0.29 \\
     $e^-e^-$        & 0.51 & 0.40 \\
     $e^-\gamma+\gamma e^-$    & 5.6 & 0.93 \\
     $e^+e^-+e^-e^+$        & 1.1 & 0.11 \\
     $e^+\gamma+\gamma e^+$     & 1.0 & 0.02 \\
        \bottomrule
    \end{tabular}
        \caption{ \label{tab:lumisummary}Total luminosity and luminosity for $\sqrt{\hat{s}}>100$~GeV for different processes at the XCC running in $\gamma\gamma$ mode at $\sqrt{s}=125$~GeV.}
\end{table}

 \begin{table}

    \centering
    \begin{tabular}{    c  | c c c r r} \toprule
     Machine   &  $E_{e^-}$ (GeV) &  Polarization & $N_{\rm H}$/yr 
     & $N_{\rm Bgnd}/N_{\rm H}$     

     &   $N_{\rm pileup/BX}$ \\ \midrule
      XCC & 62.8 & 90\% $e^-$  & 80,000  & 170  & 9.5 \\ \midrule
      OCC & 86.5 & 90\% $e^-$ & 52,000 & 1310 & 50  \\ \midrule
      ILC & 125 &  -80\% $e^-$  +30\% $e^+$ & 98,000  & 130  & 1.3 \\
      ILC & 125 & +80\% $e^-$  -30\% $e^+$ & 65,000  & 50  & 1.3 \\

        \bottomrule
    \end{tabular}
        \caption{ \label{tab:higgsratesummary} Higgs rate $N_{\rm H}$/yr and background rate $N_{\rm Bgnd}/N_{\rm H}$ for the XCC, the OCC alternative $\gamma\gamma$ collider example, and the ILC.  A year is defined as $10^7$~s, and $N_{\rm Bgnd}$ refers to the number of hadronic events with $\sqrt{\hat{s}/m_H^2}>0.64$.   The number of $\gamma\gamma\rightarrow\textrm{ hadrons}$ pileup events per bunch crossing is also indicated.}
\end{table}

\subsection{Machine Detector Interface}
Backgrounds from $e^+$, $e^-$, and $\gamma$'s produced at the Compton and primary interaction points affect the vertex detector inner radius, the beampipe radiation length, and the solid angle coverage of the main tracker/calorimeter and forward 
detectors.  Assuming a 5~Tesla solenoid field for the detector, the CAIN MC was used to simulate these backgrounds. Just as for $e^+e^-$ linear colliders, the inner radius of the vertex detector is determined by the radial extent of incoherent $e^+e^-$ pairs.  The pairs at XCC extend a little further out than at ILC, so that the vertex inner radius is set at 1.9~cm instead of 1.5~cm for ILC.

\begin{figure}
\centering
     \includegraphics[width=0.95\textwidth]{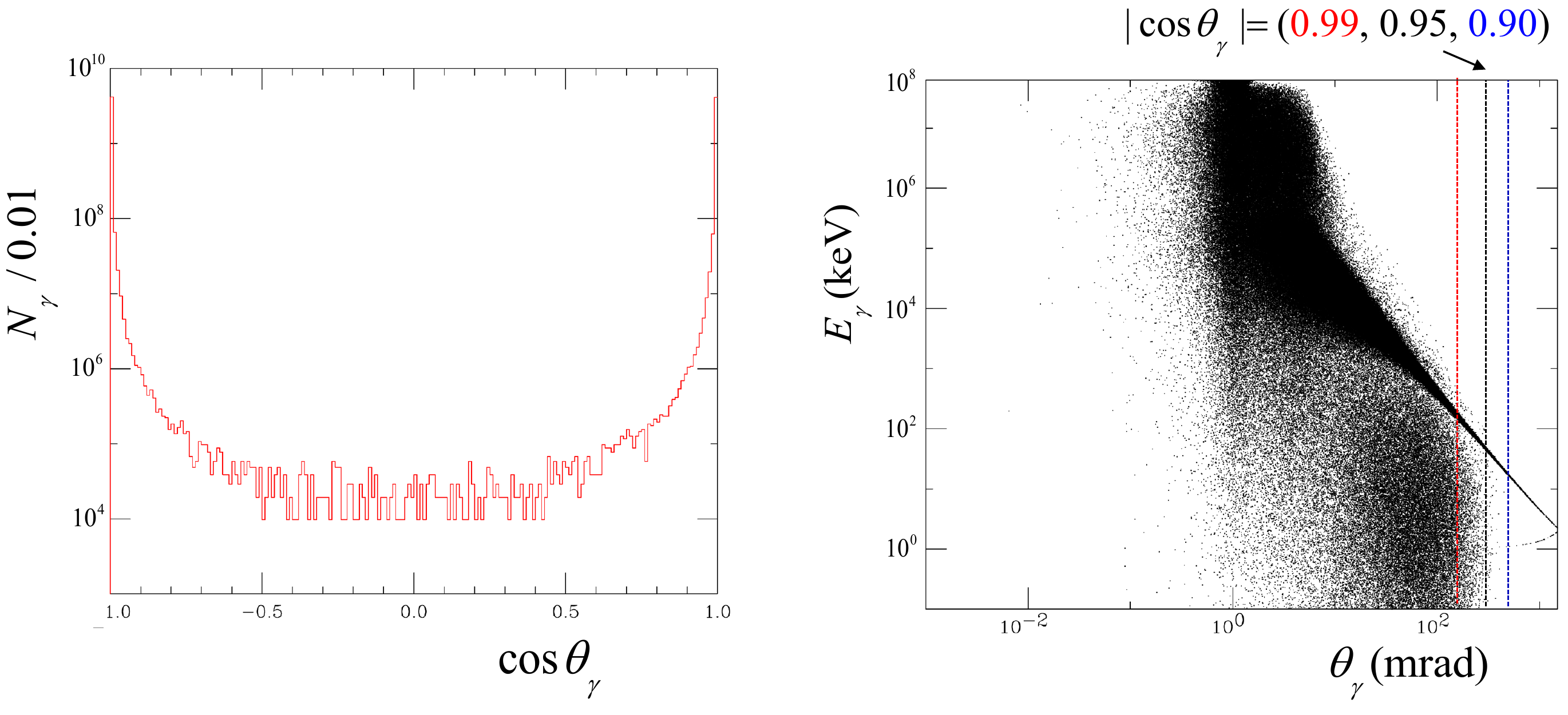}
     \caption{Number of Compton IP  photons versus $\cos{\theta_\gamma}$ (left) and photon energy versus $\theta_\gamma$ (right) where $\theta_\gamma$ is the angle of the photon
     with respect to the beam axis. 
     }
     \label{fig:xccphotons}
 \end{figure}

   Unique to XCC is the photon background from the Compton collisions, as shown in figure~\ref{fig:xccphotons}.  There is a moderate flux of soft (few keV) X-rays in the central region; these photons are dealt with by adding 0.1\% - 1.0\% $\textrm{X}_0$ of a heavy element to the beampipe for $|\cos{\theta}|<0.8$.  The number and energy of the Compton IP X-rays increases rapidly in the forward direction. The required absorber increases to 5.0\% $\textrm{X}_0$ at $|\cos{\theta}|=0.93$. A complex design may be required for $0.95<|\cos\theta|<0.99$, with limited instrumentation for $|\cos\theta|>0.99$.

\subsection{Higgs Physics at \texorpdfstring{$\sqrt{s}=125$}{Lg}~GeV\label{sec:higgsphysics}}

A comparison of the first ten years of running at ILC/C$^3$ and XCC is shown in 
table~\ref{tab:higgsfactoriesI}.  Despite a factor of 4 less luminosity, the XCC produces nearly 3 times as many Higgs bosons due to the resonance production of the Higgs.

The XCC will measure $\sigma(\gamma\gamma\rightarrow H)\times \rm{BR}(H\rightarrow X)\ \propto \Gamma_{\gamma\gamma}\Gamma_{X}/\Gamma_{\rm tot}$ in  $H$ resonance production at $\sqrt{s_{\gamma\gamma}}=125$~GeV for a variety of Higgs decay modes $X= b\bar{b},\ c\bar{c},\ WW^*$, etc.  A study with full Monte Carlo simulation of detector and background is required to understand in detail the advantages and disadvantages of XCC branching ratio measurements vis-a-vis the $e^+e^-$ colliders. Nevertheless, given the entries in table~\ref{tab:higgsratesummary}, the XCC errors for $\sigma(\gamma\gamma\rightarrow H)\times \rm{BR}(H\rightarrow X)$ should be comparable to those of ILC.    Assuming equal branching ratio precision for ILC and XCC (scaled by statistical error) the Higgs coupling precision of XCC can be calculated.
Table~\ref{tab:kappaHiggsCouplings} shows a comparison of the Higgs coupling precision of HL-LHC, ILC/C$^3$ and XCC in the $\kappa$ framework assuming no BSM decays.

 \begin{table}
    \begin{center}        
    \begin{tabular}{    l   | c c } \toprule
     & ILC/C$^3$ & XCC \\
     Colliding Particles   & $e^+e^-$ & $\gamma\gamma$ \\ \midrule
    Stage I:  & & \\ 
     \quad $\sqrt{s}$ (GeV) & 250 &  125 \\
     \quad Luminosity (fb$^{-1}$) & 2000 & 460 \\
     \quad Beam Power (MW) & 4.0 & 4.0 \\
     \quad Run Time (yr) & 10 & 10 \\ \midrule
    \# Single Higgs  & $0.5\times 10^6$ & $1.3\times 10^6$ \\
   \bottomrule
    \end{tabular}
\end{center}
        \caption{ \label{tab:higgsfactoriesI} 
        Stage I running scenario for ILC/C$^3$ and XCC including $\sqrt{s}$, luminosity, and beam power.  The number single Higgs events is also indicated. }
 \end{table}

 \begin{table}

    \begin{center}
    \begin{tabular}{    l  | c c c } \toprule
    & HL-LHC$^\dagger$ &  ILC/C$^3$ & XCC \\
     coupling $a$   &  $\Delta a$ (\%) &   $\Delta a$ (\%) &   $\Delta a$ (\%)  \\ \midrule
      $HZZ$ & 2.4 & 0.46 & 0.83 \\
      $HWW$ & 2.6 & 0.44 & 0.84  \\ 
      $Hbb$ & 6.0 &  0.83 & 0.85  \\ 
      $H\tau\tau$ & 2.8 & 0.98 & 0.89  \\ 
      $Hgg$ & 4.0 & 1.6 & 1.1 \\ 
      $Hcc$ & - & 1.8 & 1.2 \\ 
      $H\gamma\gamma$ & 2.9 & 1.1 & 0.10  \\ 
      $H\gamma Z$ & - &  -  & 1.5 \\ 
       $H\mu\mu$ & 6.7 & 4.0 & 3.5 \\ 
     
      $\Gamma_{\rm tot}$ & 5 & 1.6 & 1.7 \\
              \bottomrule
 \multicolumn{4}{l}{\footnotesize{$^\dagger$~S1 from table 36 in arXiv:1902.00134 [hep-ph]}} \\

    \end{tabular}
        
        \end{center}

            \caption{ \label{tab:kappaHiggsCouplings} Higgs coupling precision for HL-LHC and for the first 10 years of  ILC/C$^3$ and XCC, as calculated in the $\kappa$ framework under the assumptions that the Higgs only decays to Standard Model particles.The errors on 
$\sigma(\gamma\gamma\rightarrow H)\times \rm{BR}(H\rightarrow X)$ are taken from~\cite{Bambade:2019fyw}, and are assumed to be the same for ILC and XCC for all decay modes except $H\rightarrow \gamma\gamma$
and $H\rightarrow\gamma Z$, which are characterized by monochromatic photons of energy 62.5~GeV and 29.3~GeV, respectively, at XCC.}

 \end{table}

\begin{figure}
\centering
     \includegraphics[width=0.75\textwidth]{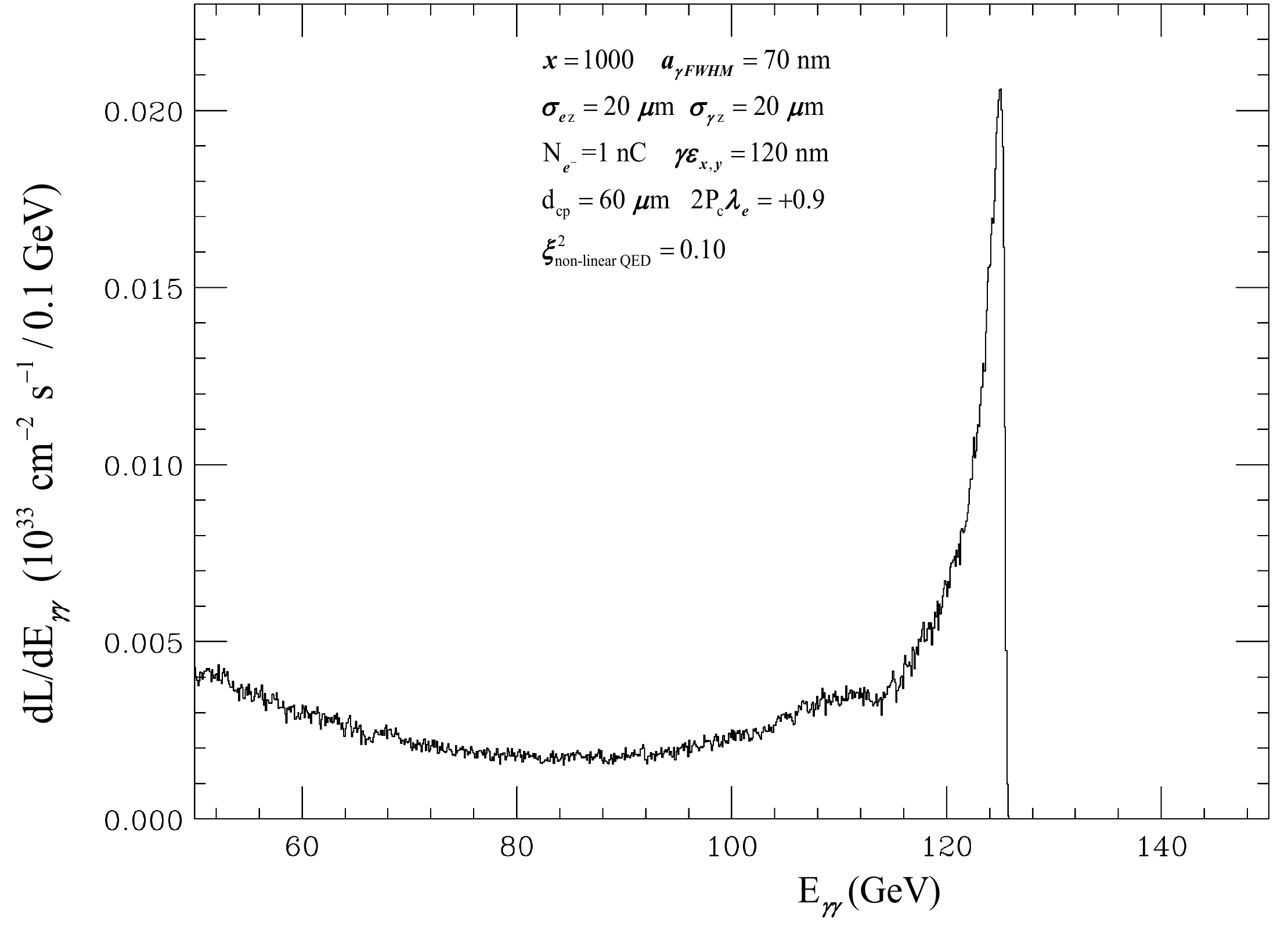}
     \includegraphics[width=0.75\textwidth]{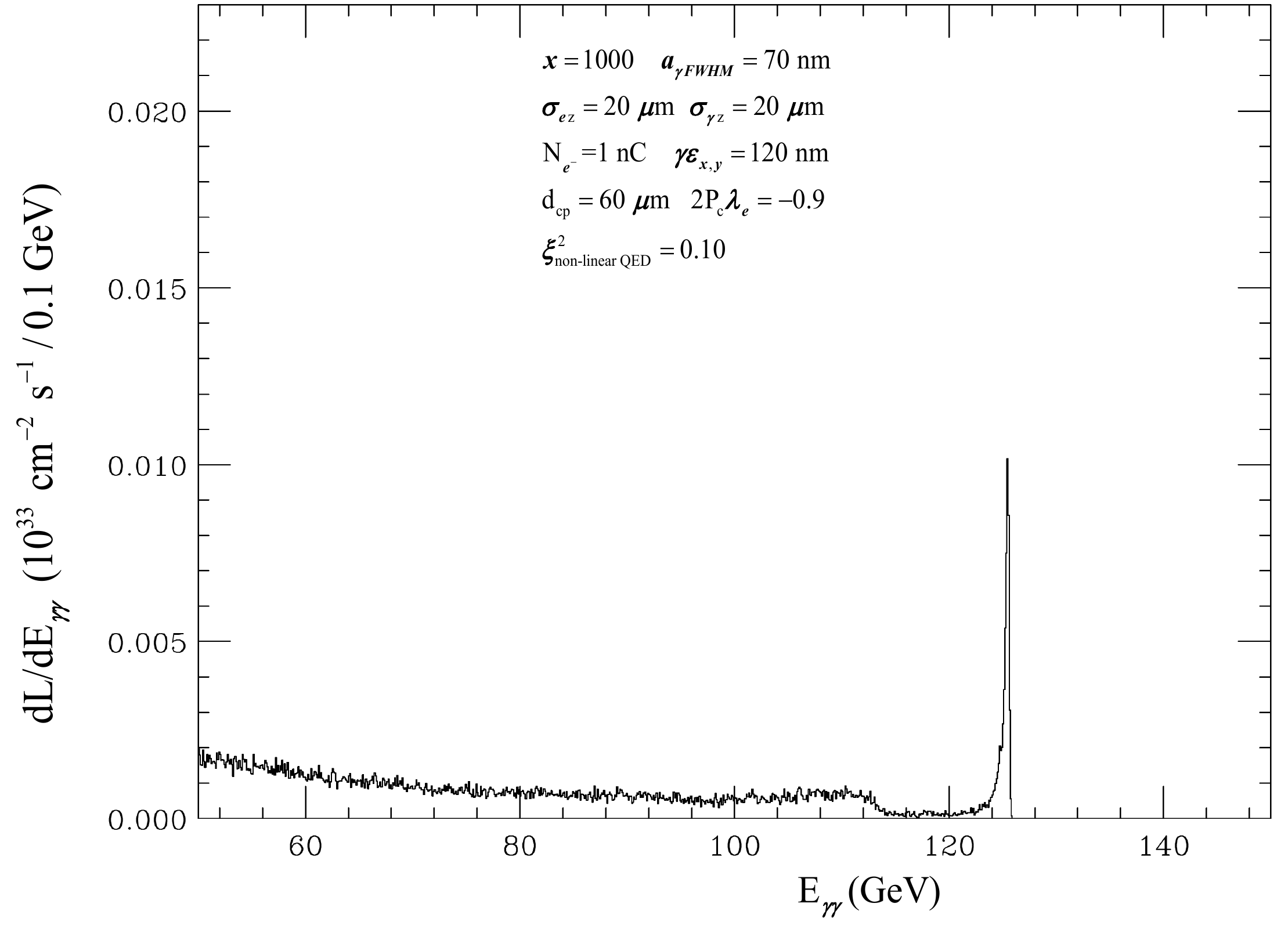}
     \caption{$\gamma\gamma$ luminosity as a function of $\gamma\gamma$ center-of-mass energy $E_{\gamma\gamma}$ for x=1000 \& $2P_c\lambda_e=+0.9$ (top) versus $2P_c\lambda_e=-0.9$ (bottom) as calculated by the  CAIN MC, where $P_c$ and $\lambda_e$ are the helicities of the laser photon and electron, respectively.  The x-ray waist radius $a_\gamma$ at the Compton interaction point and the electron (x-ray) beam r.m.s longitudinal sizes, $\sigma_{ez}$ ($\sigma_{\gamma z}$), are indicated. The 0.10\% electron beam energy spread, linear QED Bethe-Heitler scattering and non-linear QED effects in Compton and Breit-Wheeler scattering are included in the CAIN simulation.
     }
     \label{fig:x1000lumiminus}
 \end{figure}

The total Higgs width $\Gamma_{\rm tot}$ can be measured with a scan of the XCC energy across the Higgs resonance. An alternate XCC polarization $2P_c\lambda_e=-0.9$ provides
a narrower leading edge width of 0.045~GeV (dominated by the 0.1\% $e^-$ beam energy spread) as shown in figure~\ref{fig:x1000lumiminus}. This comes at a cost, however, as the Higgs rate is reduced to 40,000 Higgs per year.  With such a configuration an energy scan can be used to measure the total Higgs width to an accuracy of $\Delta \Gamma_{\rm tot}=4.5$~MeV.  This is fine if $\Gamma_{\rm tot}$ is a few 10's of MeV, but clearly insufficient in the more likely event that the true total Higgs width is close to its Standard Model value $\Gamma_{\rm tot}\approx4.0$~MeV.

The $\gamma\gamma$ partial width $\Gamma_{\gamma\gamma}$ can be measured directly in the process $e^-\gamma\rightarrow e^- H$ if $\sqrt{s} > 125$~GeV. Such a quantity can be used to convert the XCC branching ratios into  model independent coupling measurements. This will be discussed in detail in the $\sqrt{s}=380$~GeV energy upgrade section.

\subsection{Cost: XCC vs. \texorpdfstring{$e^+e^-$}{Lg} Higgs Factories}

Using the C$^3$-250 cost model\cite{Bai:2021rdg}, an initial estimate of the XCC cost breakout is given in table~\ref{tab:capitalcost}.   When compared to the capital cost of \$3.7B for one specific C$^3$-250 scenario in~\cite{Bai:2021rdg} the
 XCC represents a savings of nearly 40\%.  Given the very early stage of the XCC design it is important that table~\ref{tab:capitalcost} be viewed as illustrative, providing insight into the potential cost savings of the XCC.

 \begin{table}

    \centering
    \begin{tabular}{    c  | c | c | c} \toprule
    & Sub-Domain & \%  & \% \\ \midrule
    Sources                & Injectors                      & 9  & 26 \\
                           & FEL                            & 9  &    \\
                           & Beam Transport                 & 9  &    \\ \midrule
 Main Linac                & Cryomodule                     & 9  & 30   \\
                           &  C-band Klystron               & 22  &     \\  \midrule
    BDS                    &  Beam Delivery and Final Focus &  7  &  15  \\
                           &         IR                     &  8  &    \\ \midrule
Support Infrastructure     & Civil Engineering              &  5  &  28  \\
                           &  Common Facilities             & 18  &    \\
                           &  Cryo-plant                    &  6  &    \\ \midrule 
        Total              &      2.3B\$                    & 100 & 100   \\

        \bottomrule
    \end{tabular}
        \caption{ \label{tab:capitalcost} Initial estimate of  XCC cost breakout using the C$^3$-250 cost model.}
 \end{table}

\section{Design Overview}
\subsection{Attainable Energy}
\subsubsection{Electron Accelerator} \label{accelerator}

The C$^3$ technology represents a new methodology for dramatically reducing the cost of high gradient accelerators, while increasing their capabilities in terms of gradient and efficiency. After two decades of exploring the high gradient phenomena observed in room-temperature accelerator structures, the underlying physics models related to these phenomena have been deduced. This knowledge led  to the creation of a new paradigm for the design of accelerator structures, which includes: a new topology for the structure geometry \cite{Sami1,Sami2} operating at cryogenic temperature, the use of doped copper in the construction of these structures \cite{Sami4}, and a new methodology for the selection of operating frequency bands \cite{Sami4}. In particular, for science discovery machines, optimization exercises have revealed that the optimal frequency should be around 6--8 GHz for operation with a gradient well above 100 MeV/m while maintaining exquisite beam parameters.  That explains why both UCLA and LANL are trying very hard to build their infrastructure at C-band (5.712 GHz), a frequency band that is close enough to the optimal point, but with some industrial support behind it. 

Furthermore, the so-called ``distributed-coupling structure" \cite{Sami1} and its operation at cryogenic temperature represent a breakthrough for the e$^-$ source. Electron guns can be designed around this concept with an unprecedented brightness  \cite{Sami6}. Using this technology can result in an extremely economical system for this $\gamma$-$\gamma$ collider. The two Linacs required for the collider could be made extremely compact due to the high gradient capabilities of the C$^3$ technology and the limited energy reach required of 62.5 GeV. With the bright electron beam sources, damping rings can be eliminated.   The Linac parameter set is included in tables~\ref{tab:design} and~\ref{tab:design380}.  A description of C$^3$ technology applied to an $e^+e^-$ collider can be found here\cite{Bai:2021rdg}.

\subsubsection{X-ray FEL }

The two identical X-ray FEL lines, which provide the necessary circularly-polarized 1.2 nm (1 keV) photons, can be constructed using a long helical undulator. A full time-dependent GENESIS study (described below) has validated the high magnetic field and high electron energy design considered here.  The quantum diffusion energy spread in such an undulator must be taken into account, and will be properly included in future studies. As the main Linac can accelerate electrons to 62.5 GeV, the electron energy for the XFEL line is taken to be around 31 GeV, with normalized emittance of 120 nm, bunch charge of 1 nC and relative RMS slice energy spread of $\langle\Delta\gamma/\gamma\rangle$ of 0.05\%.

\begin{figure}
\centering
     \includegraphics[width=0.75\textwidth]{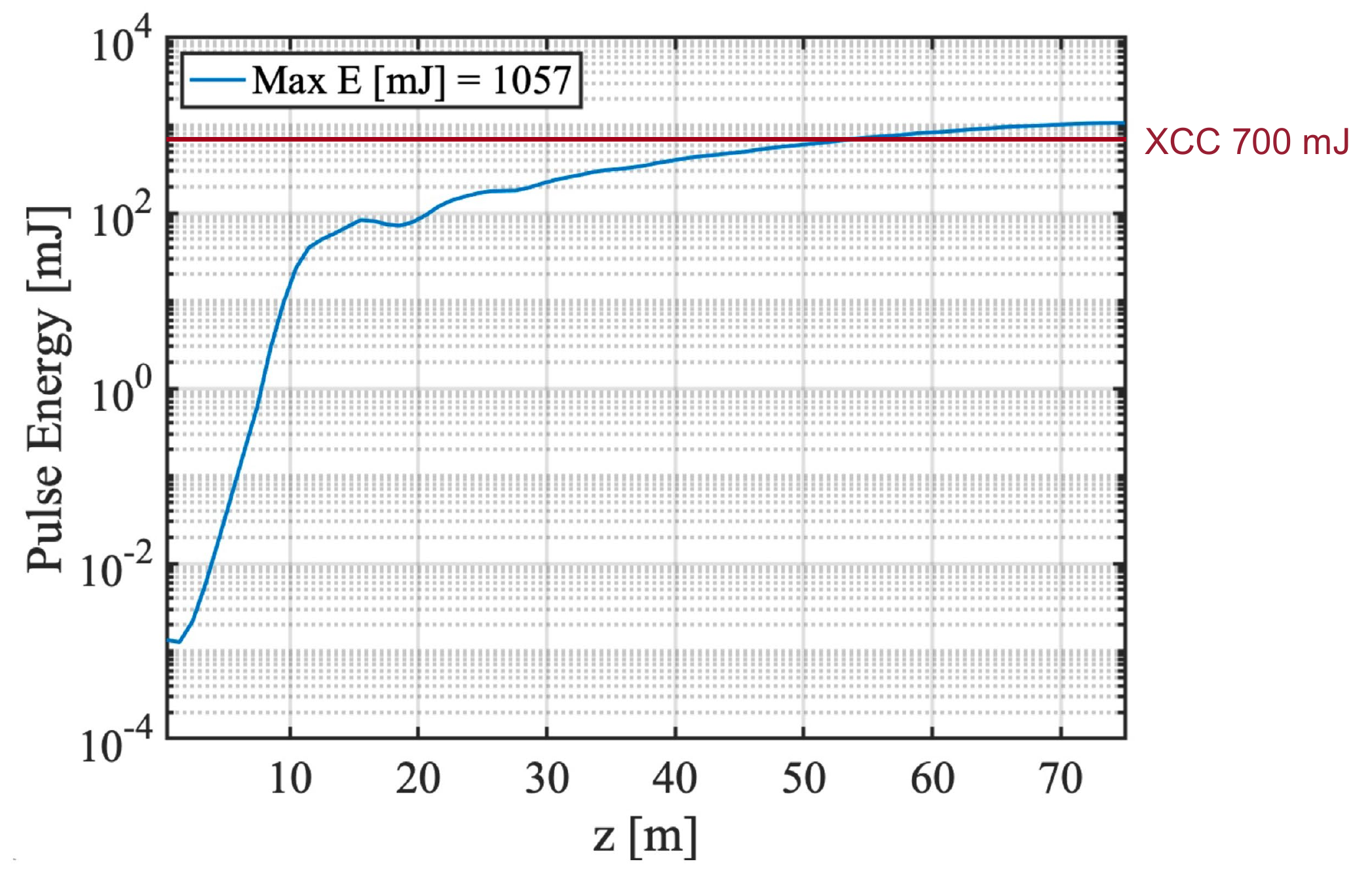}
     \caption{GENESIS simulation of X-ray pulse energy versus distance along the XCC XFEL undulator.}
     \label{fig:pulseevsz}
 \end{figure}

Using a permanent-magnet undulator, with peak magnetic field slightly above 1 Tesla, undulator period around 9 cm and an average $\beta$-function of 12 m, a  1~keV X-ray beam with pulse energy $\sim 0.07$~J at FEL saturation length of roughly 60 m and with negligible quantum diffusion effects~\cite{KimHuangBook,HuangReviewPaper} can be produced. As is known from a decade of X-ray FEL studies, if a seeded FEL (such as through self-seeding or other similar processes) is produced and the undulator's $K$ parameter after saturation is tapered, then X-ray pulse energy can continue to be extracted with an order-of-magnitude improvement in efficiency ~\cite{TaperPaper}. The targeted pulse energy of 0.7 J at 1 keV photon energy can then be reached, which is about 2.3\% of the electron beam energy. The overall length of the undulators is estimated to be within 200 m. This is just an example parameter set (summarized in table~\ref{tab:design} below). 

The XFEL design has been simulated with GENESIS 1.3. The simulation is run with parameters similar to those shown in table 2. The difference between those and the simulated parameters are the initial electron slice energy spread (0.01$\%$), electron energy (30 GeV), peak current (6kA), average beta function (2 m), undulator period (5 cm) and undulator peak field (1.8 T). The undulator is a helical permanent magnet undulator with a super-imposed alternating gradient focusing quadrupole lattice. The undulator is split into two sections, the first one generates Self Amplified Spontaneous Emission (SASE) \cite{Bonifacio1984} and the second is a self-seeded section in which a monochromatic seed is overlapped with the electron beam after passing through an idealized monochromator. The time dependent SASE simulation produces a 0.4 mJ and $\sim$ 100 fs FWHM X-ray pulse after a 15 m undulator. The self-seeded section assumes 0.3 $\%$ of the SASE X-ray power is filtered through the monochromator and a quadratic post-saturation taper is applied to increase the pulse energy following the exponential gain region. As shown in Fig.~\ref{fig:pulseevsz}, the resulting X-ray energy after a 75 m undulator is 1.05 J with a 3.5 $\%$ extraction efficiency, in reasonable agreement with the analytic estimates obtained using the parameters of table 2. It should be noted that a similar extraction efficiency has already been achieved experimentally for tapered self-seeded systems XFELs \cite{Emma2017}.


\begin{table}[htbp]
\begin{center}
\caption{\label{tab:design} Summary of design parameters for $\sqrt{s}=125$~GeV.}
\begin{tabular}{ |l|l|||l|l|c| }
\hline
Final Focus parameters  & Approx.~value & XFEL parameters & Approx.~value \\
\hline
\hline
Electron energy & 62.8 GeV & Electron energy & 31 GeV \\ 
Electron beam power & 1.24 MW & Electron beam power &  0.61 MW \\
$\beta_x/\beta_y$ & 0.030/0.030 mm & Normalized emittance & 120 nm \\ 
$\gamma\epsilon_x/\gamma\epsilon_y$ & 120/120 nm & RMS energy spread $\langle\Delta\gamma/\gamma\rangle$ &  0.05\% \\ 
$\sigma_x/\sigma_y$ at $e^-e^-$ IP & 5.4/5.4 nm & Bunch charge & 1 nC \\ 
$\sigma_z$ & 20 $\mu\rm{m}$  & Linac-to-XFEL curvature radius & 133 km  \\ 
Bunch charge & 1 nC &  Undulator B field & $\gtrsim$ 1 T \\ 
Bunches/train at IP & 165 & Undulator period $\lambda_u$ &  9 cm \\ 
Train Rep. Rate at IP & 120 Hz & Average $\beta$ function &  12 m \\
Bunch spacing at IP & 4.2 ns &  x-ray $\lambda$ (energy) &  1.2 nm (1 keV) \\
$\sigma_x/\sigma_y$ at IPC & 12.1/12.1 nm & x-ray pulse energy &  0.7 J  \\
$\mathcal{L}_\textrm{geometric}$ & $2.1\times 10^{35}\ \textrm{cm}^2\ \textrm{s}^{-1}$ & rms pulse length &  20 $\mu\rm{m}$  \\
$\delta_E/E$ & 0.1\% & $a_{\gamma x}$/$a_{\gamma y}$ (x/y waist) & 21.2/21.2~nm\\
$L^*$ (QD0 exit to $e^-$ IP) &  1.5m  & non-linear QED $\xi^2$ & 0.10  \\ 
$d_{cp}$ (IPC to IP) & $60\ \mu$m  & & \\
QD0 aperture & 12 cm diameter & & \\
Accel. gradient & 70 MV/m & & \\  \midrule
Site parameters & Approx. value & & \\ \midrule
crossing angle & 2 mrad & & \\
total site power & 115 MW & & \\
total length & 4.2 km & & \\
\hline
\end{tabular}
\end{center}
\end{table}

\subsection{Attainable Luminosity}
\subsubsection{Electron Final Focus}

\begin{figure}
\centering
    \includegraphics[width=0.95\textwidth]{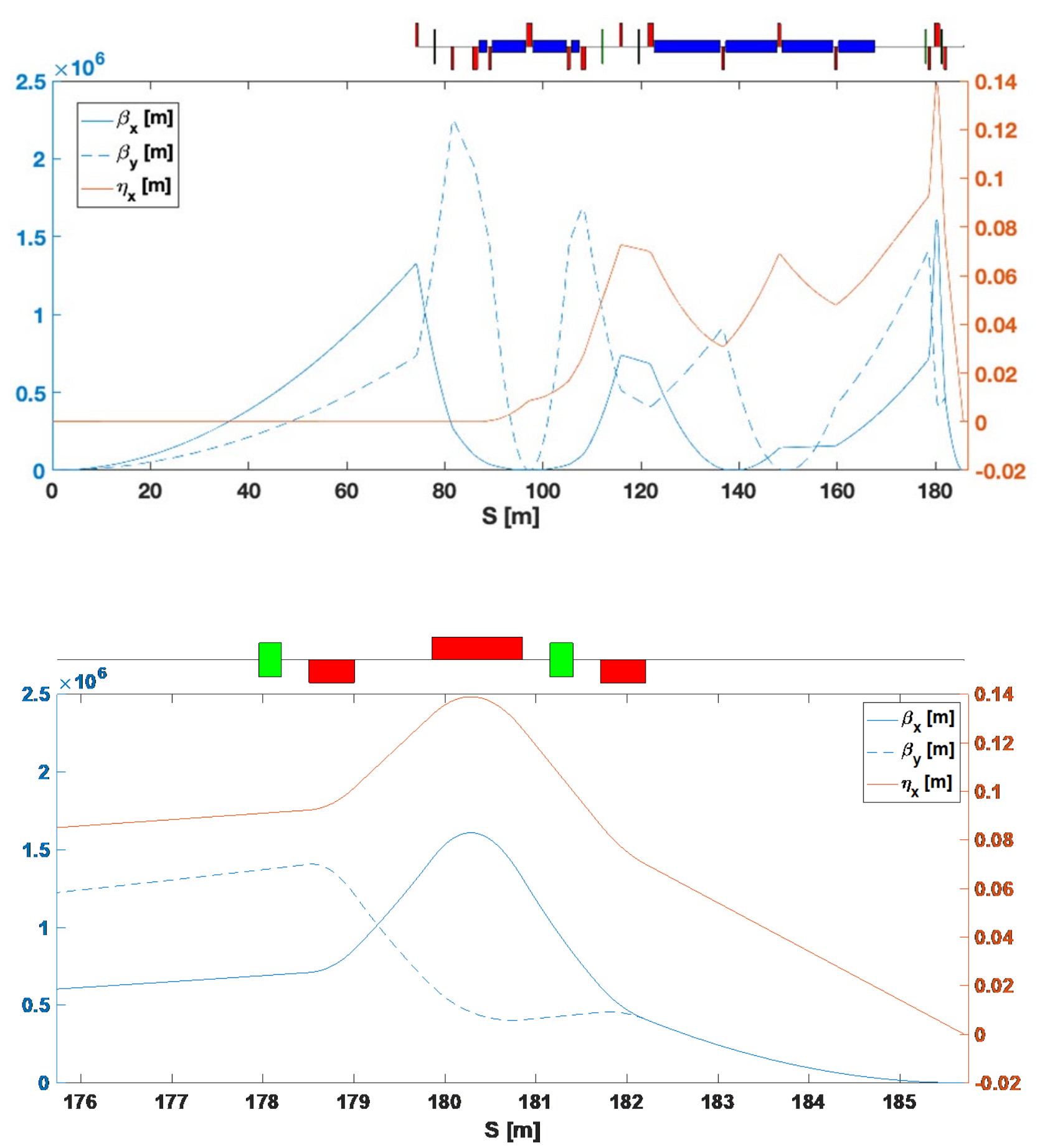}    
     \caption{Beta and horizontal dispersion functions for final focus beamline (IP is on the right). The bottom figure shows a close up of the IP triplet.}
     \label{fig:betadisp}
 \end{figure}

A preliminary layout for the final focus system is shown in Fig.~\ref{fig:betadisp}.
This is not an optimized design at this stage and is shown for illustration purposes only. 
The length of the system is about 110m as shown, using realistic magnet strengths. The design follows 
that used for ILC and CLIC, namely the local chromatic compensation scheme proposed by Raimondi \& Seryi\cite{Raimondi:2000cx} and tested at ATF2, KEK\cite{White:2014vwa}. The design uses a pair 
of sextupole magnets located locally to the final triplet to cancel the chromaticity generated by the final focus system magnets. An interleaved, second pair of sextupoles are 
used to simultaneously cancel geometric aberrations introduced by the chromatic correction sextupoles, with a fifth sextupole 
used to help control third-order aberrations generated by the interleaved sextupole pairs. Octupoles, decupoles (and perhaps higher harmonic magnets) will also be required but are not 
shown here. Horizontal dispersion, required for the chromatic cancelation to occur, is generated by three families of 
bend magnets. The length of the bend magnets is chosen such 
that negligible emittance growth due to synchrotron radiation exists.

A crab cavity system rotates the electron bunches so that they collide head on despite the $\theta_c=2$~mrad crossing angle.  Since the XFEL photons exchange their momenta with the electrons in the Compton process, the loss of $\gamma\gamma$ luminosity due to the crossing angle will also be recovered with the crab cavity scheme (the collision of the XFEL photons with the rotated electron bunches is simulated by the CAIN MC).  The XCC  Piwinski angles $\Phi_p=\theta_c \sigma_z/2\sigma_x=3.7$ (7.7) radians for $\sqrt{s}=125$~GeV (380~GeV) are similar to the angles $\Phi_p=4.1$ and 7.3 radians for ILC at $\sqrt{s}=250$~GeV and CLIC at $\sqrt{s}=1500$~GeV, respectively.
 \subsubsection{X-ray Optics}\label{xrayoptics}
 The x-ray beam must be focused from a waist radius of $a_\gamma=9000$~nm at the undulator exit to $a_\gamma=30$~nm (70~nm FWHM) at the Compton interaction point (IPC) in order to match the  transverse size of the electron beam at the IPC.  The distance $d_{cp}=60\ \mu$m between the IPC and IP cannot be made much larger due to the angular spread of the converted 62.5 GeV photons.   The angle $\theta$
 that a converted photon makes with respect to the electron direction is correlated with its energy $\omega$ via $\theta=\theta_0\sqrt{\omega_m/\omega-1}$, where $\omega_m=62.8$~GeV is the maximum photon energy and $\theta_0=\sqrt{x+1}/\gamma_e=0.26$~mrad.  A photon 
 with energy $m_H/2=62.5$~GeV therefore makes an angle $\theta=0.018$~mrad,
 so that the design value $d_{cp}=60\ \mu$m
 results in a 1.1~nm increase in the transverse size of the 62.5~GeV photon beam.  This should be compared to the 5.4~nm transverse size of the photon beam at the IP assuming no Compton angular spread.
 As an example of an alternate configuration, if the distance between the IPC and IP  were increased to $d_{cp}$=100~nm, the $\gamma\gamma$ luminosity would drop by 30\% with respect to the luminosity with $d_{cp}=60\ \mu$m.

\begin{figure}
\centering
     \includegraphics[width=0.85\textwidth]{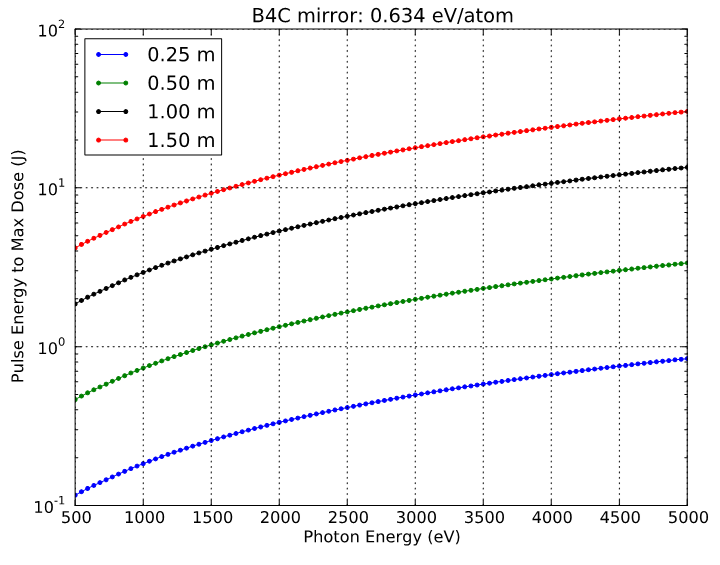}
     \caption{Single pulse mirror damage limit versus photon energy for KB mirror substrate lengths of 0.25, 0.50, 1.00 and 1.50 meters. }
     \label{fig:MirrorDamageLimitSinglePulse}
 \end{figure}

\begin{figure}
\centering
     \includegraphics[width=0.85\textwidth]{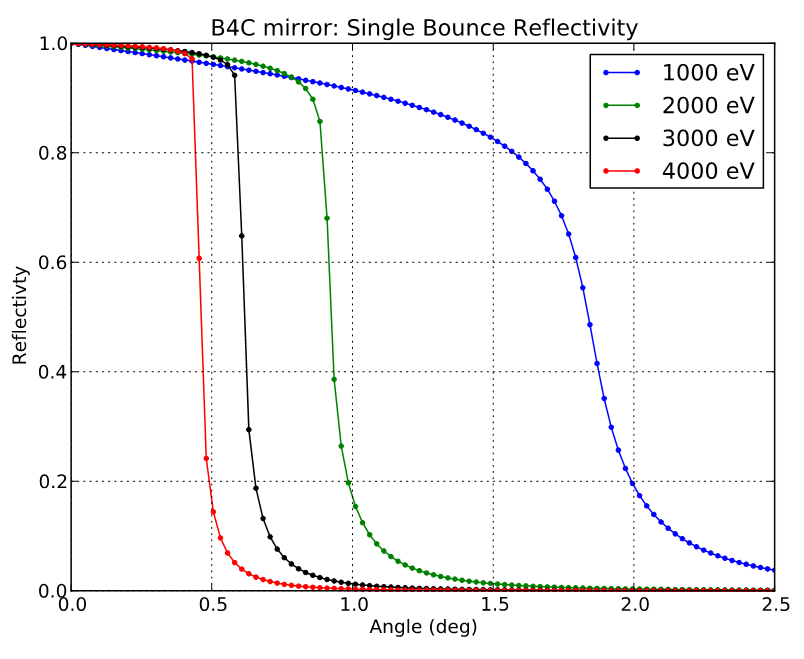}
     \caption{Mirror reflectivity versus angle of incidence for different photon energies}
     \label{fig:MirrorRefelectivity}
 \end{figure}




\begin{figure}
\centering
     \includegraphics[width=0.85\textwidth]{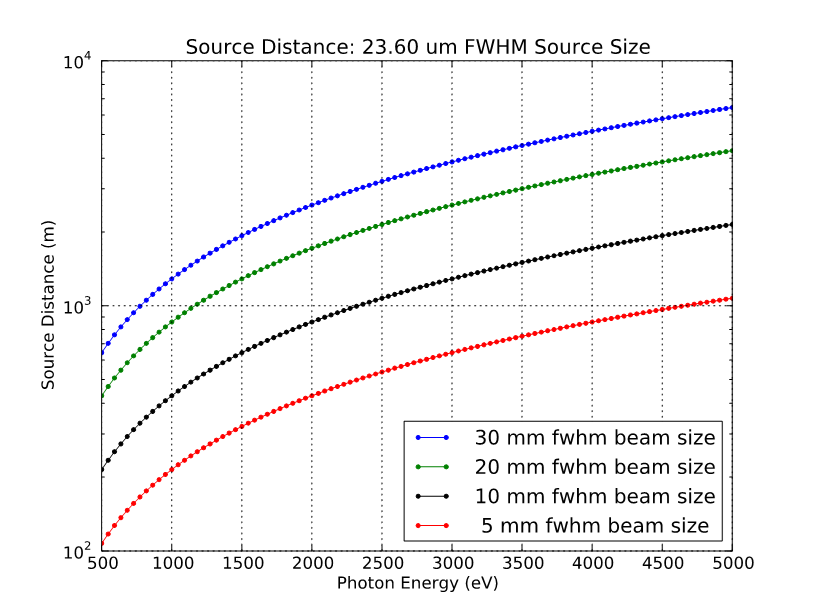}
     \caption{Source distance --  the distance between the downstream edge of the XFEL and the upstream edge of the KB mirror module -- versus photon energy for different mirror-incident beam sizes.}
     \label{fig:SourceDistance}
 \end{figure}

\begin{sidewaysfigure}
\centering
     \includegraphics[scale=0.7]{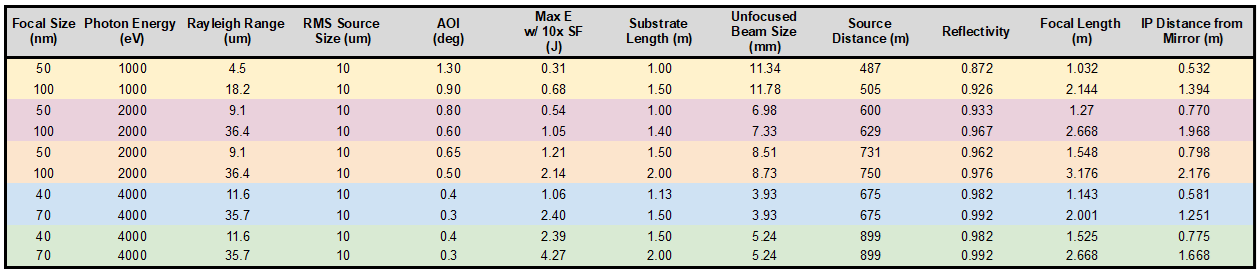}
     \caption{X-ray Optics Parameter Table}
     \label{fig:X-rayOpticsParameterTable}
 \end{sidewaysfigure}

An X-ray optics system for the XCC has not yet been designed. Nevertheless,
 an initial X-ray optics study has been performed to determine broadly the design parameters required  to focus 1~Joule/pulse soft x-rays to 70~nm FWHM focal size using Kirkpatrick-Baez (KB) mirrors. Boron carbide ($\textrm{B}_4\textrm{C}$) is the highest damage threshold coating and is used in the study. Fig.~\ref{fig:MirrorDamageLimitSinglePulse} shows the single pulse energy damage limit as a function of photon energy for KB mirror substrate lengths of 0.25, 0.50, 1.00 and 1.50 meters. The calculation is only weakly dependent on the incident angle below cutoff;  an angle of incidence of 0.3$^\circ$ was used here.  A safety factor of five to ten times these limits should be included in the actual design.  The general trend is that longer substrates and higher photon energies are better.  The XCC XFEL photon energy could perhaps be pushed up from 1~keV to 2~keV to accommodate X-ray focusing issues. However, going much beyond 2~keV would result in too great a loss in Higgs rate due to the falling Compton cross section with higher photon energies.  The lesson from Fig.~\ref{fig:MirrorDamageLimitSinglePulse} is that long ($>1$~meter) KB mirrors are needed to focus $\sim 1$~Joule/pulse soft X-ray beams.

 Fig.~\ref{fig:MirrorRefelectivity} shows mirror reflectivity versus angle of incidence for photon energies of 1,2,3, and 4~keV.  Again, the trend is that higher photon energies are better as they have higher mirror reflectivity and so present a  smaller heat load for the KB mirror cooling system.  Fig.~\ref{fig:SourceDistance} contains the dependence of the source distance (i.e., the distance between the downstream edge of the XFEL and the upstream edge of the KB mirror module)  versus photon energy for mirror-incident (unfocused) beam sizes of 5, 10, 20 and 30~mm FWHM.  The focal length will be longer for larger incident beam sizes.  A long focal length is required to simultaneously maintain an $e^-\gamma$ collision angle of less than 2~mrad and avoid congestion near the interaction point where, for example, the downstream edge of the 12~cm aperture final focus quad is located 1.5 meters from the IP.  Long source distances of order 1~km appear to be desirable, but have not yet been incorporated into the XCC design.

 Based on these consideration, Fig.~\ref{fig:X-rayOpticsParameterTable} contains a table of possible X-ray optics design parameters.  Since the horizontal and vertical KB mirrors are sequential,  a round incoming source will be focused to an ellipse. The focal size therefore refers to a round beam equivalent of $\sqrt{a_x  a_y}$ where $a_x$ and $a_y$ are the horizontal and vertical spot sizes, respectively. With a factor of ten safety factor for mirror damage, the two rows corresponding to 
 1~keV don't quite provide a solution for 0.72~J pulse energy.  Solutions are found for 2~keV photons, especially for very long substrates of 1.5~m or 2.0~m. 

 In summary, long ($>1$~m) mirrors are needed for 1~Joule pulse energies.  It should be noted that 1~m FEL quality substrates are produced today, and 1.5~m substrates have been produced for synchrotrons.  FEL-quality substrates with lengths between 1~m and 2~m would require development with industry but not R\&D.    Substrate lengths greater than 2~m are beyond state-of-the-art, and should be pursued. Finally, it could  be that X-ray optics considerations push the XCC design photon energy up from 1~keV to 2~keV.

 \subsubsection{Beam-beam Effects and Luminosity Integral}
 
 The CAIN Monte Carlo program is used to simulate beam-beam effects at both the IPC and IP. For  laser scattering at the IPC, the  CAIN program includes non-linear QED effects in Compton scattering ($e^-\gamma_0\rightarrow e^-\gamma$) and Breit-Wheeler scattering ($\gamma\gamma_0\rightarrow e^+e^-$) where $\gamma$ and $\gamma_0$ refer to the Compton-scattered and laser photon, respectively.  Bethe-Heitler scattering with laser photons ($e^-\gamma_0\rightarrow e^-e^+e^-$) is not included in CAIN, and must be added to the code.  To include this process, the equivalent photon approximation is used for virtual photons $\gamma^*$ radiated by the initial state electron,
 and the code for $\gamma\gamma_0\rightarrow e^+e^-$ is then used to simulate $\gamma^*\gamma_0\rightarrow e^+e^-$.  Non-linear QED Bethe-Heitler scattering with laser photons is not simulated.

 In the nominal XCC polarization configuration of $2P_c\lambda_e=+0.9$, the electron and laser photon helicities are given the same sign, which leads to collision lengths of 34, 25, and 95~$\mu\rm{m}$ for the Compton process
$e^-\gamma_0\rightarrow e^-\gamma$, the trident process $e^-\gamma_0\rightarrow e^-e^+e^-$,
and the  $\gamma\gamma_0$ annihilation process $\gamma\gamma_0\rightarrow e^+e^-$, respectively.      The $\gamma\gamma_0$ annihilation collision length is $3\times$  longer than it would have been if the electron and laser beams were unpolarized, and
$5\times$ longer than the collision length if the electron and laser beams had been given opposite helicities. With a collision length of 6.3 times the total laser pulse length, the $\gamma\gamma_0$ annihilation process is a nonissue.
The total conversion efficiency of electrons to primary first generation photons is 18\%. 

 Beamstrahlung and coherent pair-production at the IP are fully simulated by the CAIN MC. The beamstrahlung is significant, as can be seen in the $\gamma\gamma$ luminosity distribution for $E_{\gamma\gamma}>100$~GeV in figure~\ref{fig:x1000lumiplus}.  However, the effective IP beamstrahlung parameter $\Upsilon=5r_e^2\gamma N/6\alpha\sigma_z(\sigma_x+\sigma_y)$ is appreciatively smaller than the naive value $\Upsilon=3.2$ calculated using the unperturbed electron IP beam size $\sigma_x=\sigma_y=5.4$~nm from table~\ref{tab:design}.  Following Compton scattering at the IPC, the r.m.s. electron beam size is $\sigma_x=\sigma_y=37$~nm at the start of the IP collision, increasing to  $\sigma_x=\sigma_y=700$~nm midway through.  The beamstrahlung parameter therefore varies within the range $0.01<\Upsilon<0.5$ throughout the collision with a mean value $\langle\Upsilon\rangle=0.074$.

\begin{figure}
\centering
     \includegraphics[width=0.75\textwidth]{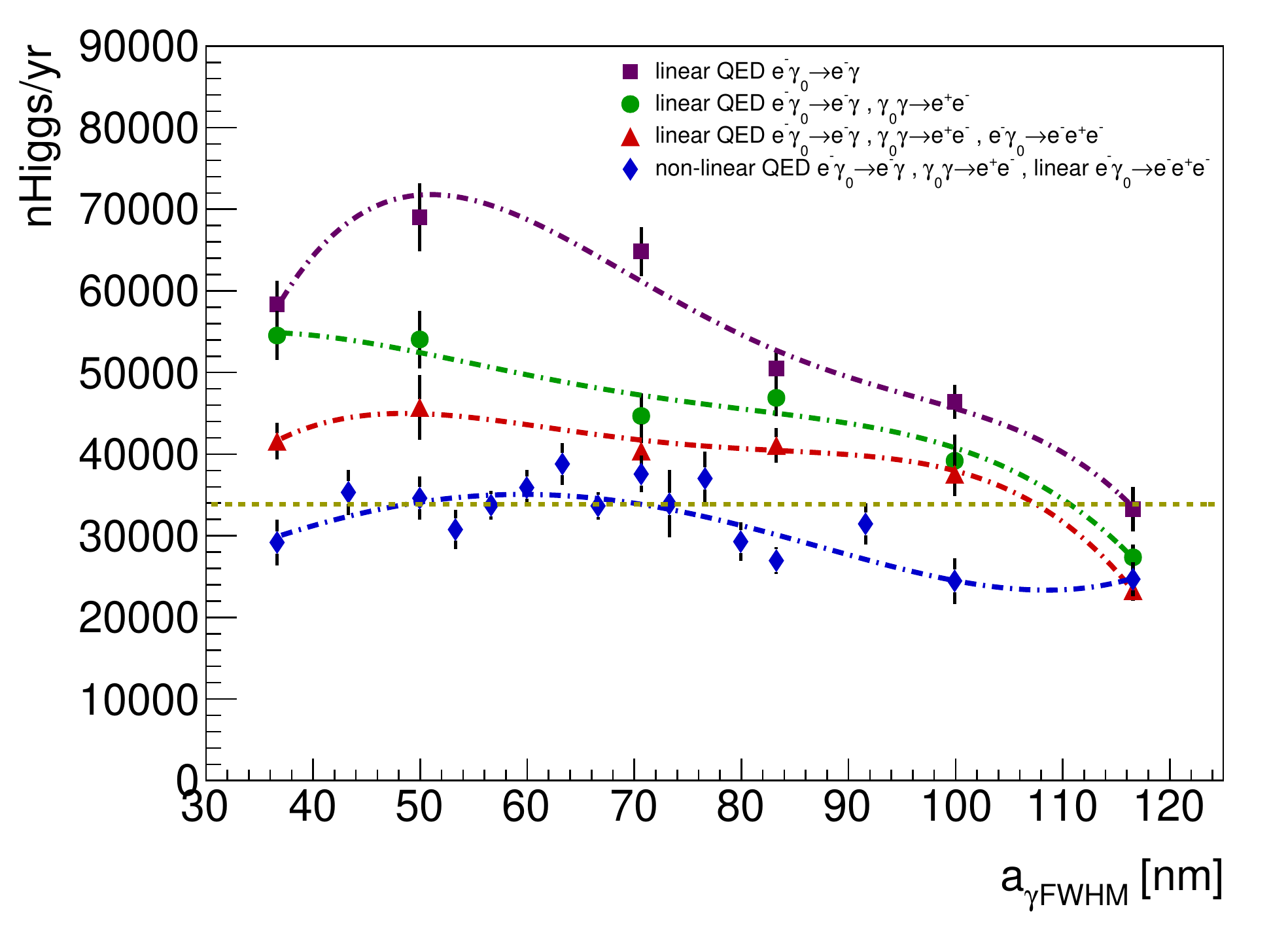}
     \caption{Scans of the FWHM spot size ${\rm a}_{\gamma FWHM}$ of the XFEL beam at the Compton interaction point (IPC) showing the impact of different IPC processes on the Higgs production rate assuming 1.7~MW total beam power. In the expressions for the IPC processes the symbols 
     $\gamma_0$ and $\gamma$ refer to laser photon and scattered photon, respectively. 4th order polynomial fits to the CAIN simulation results are indicated by dashed lines. The gold horizontal dashed line corresponds to the XCC design Higgs production rate.}
     \label{fig:lumVsAgamma}
 \end{figure}

\begin{figure}
\centering
     \includegraphics[width=0.75\textwidth]{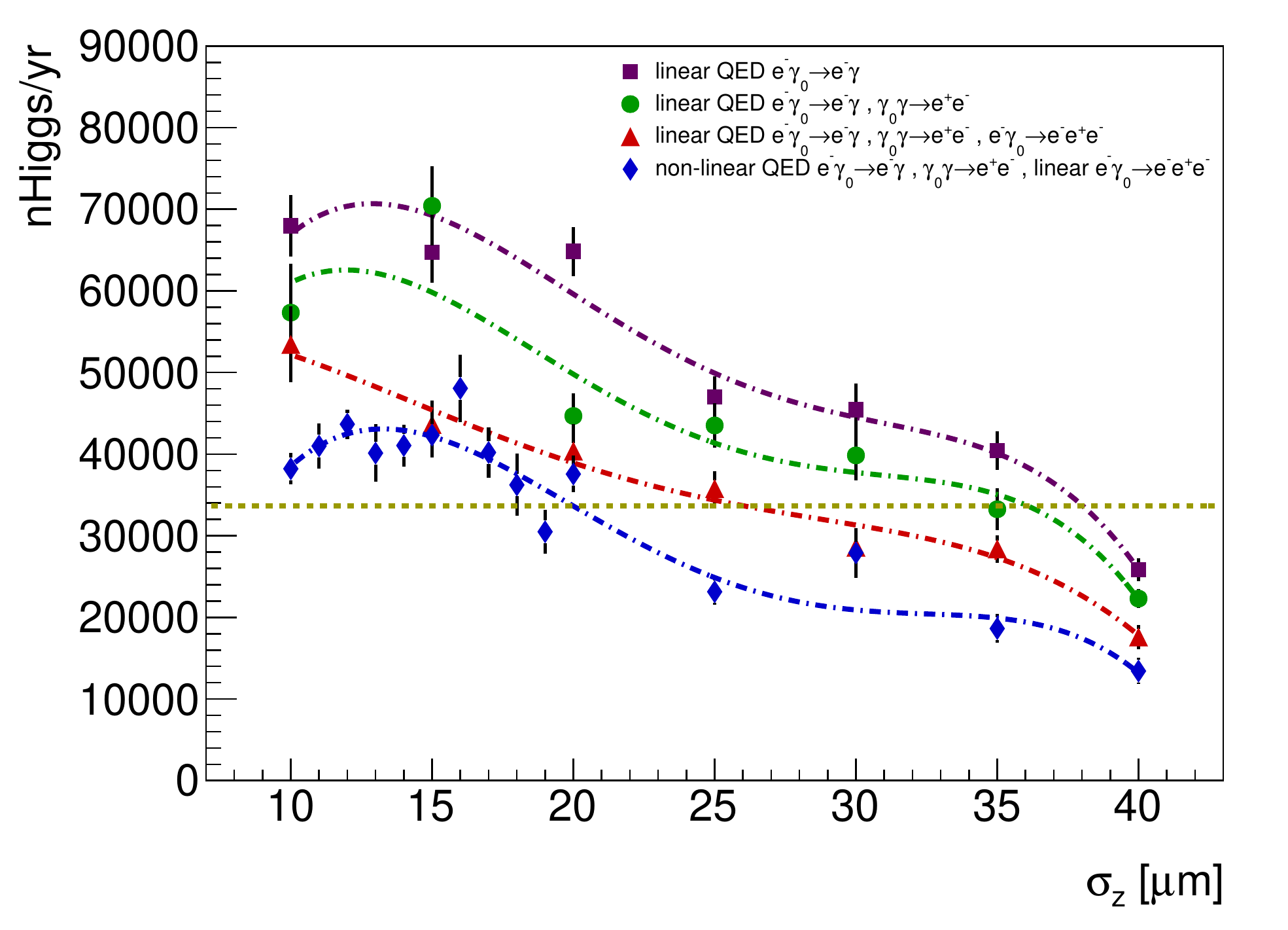}
     \caption{Scans of the r.m.s. electron longitudinal  bunch density $\sigma_{ez}$ showing the impact of different IPC processes on the Higgs production rate assuming 1.7~MW total beam power. In the expressions for the IPC processes the symbols 
     $\gamma_0$ and $\gamma$ refer to laser photon and scattered photon, respectively.  4th order polynomial fits to the CAIN simulation results are indicated by dashed lines. The gold horizontal dashed line corresponds to the XCC design Higgs production rate.}
     \label{fig:lumVsSigez}
 \end{figure}

\begin{figure}
\centering
     \includegraphics[width=0.75\textwidth]{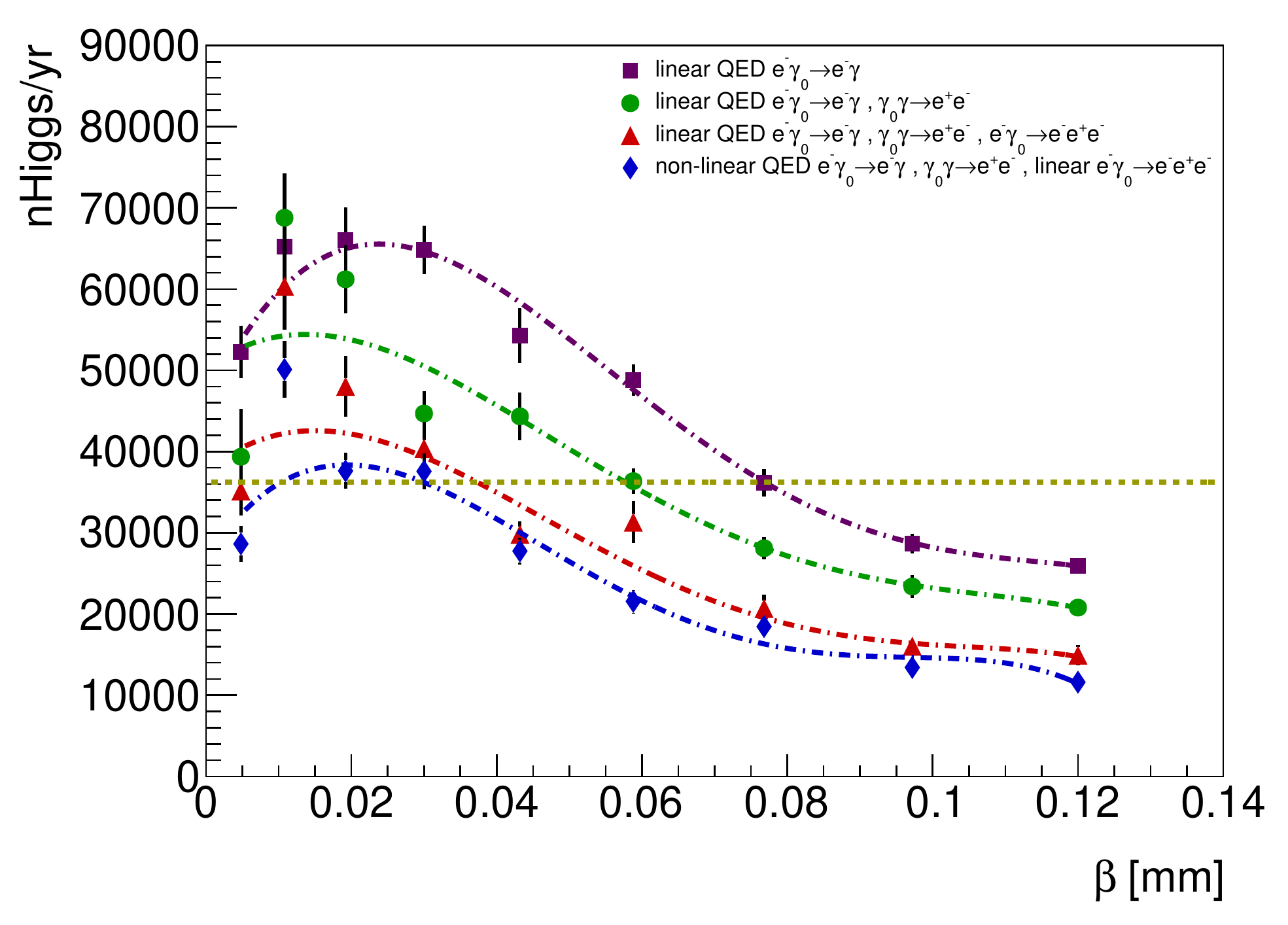}
     \caption{Scans of the electron final focus $\beta$ value showing the impact of different IPC processes on the Higgs production rate assuming 1.7~MW total beam power. In the expressions for the IPC processes the symbols 
     $\gamma_0$ and $\gamma$ refer to laser photon and scattered photon, respectively.  4th order polynomial fits to the CAIN simulation results are indicated by dashed lines. The gold horizontal dashed line corresponds to the XCC design Higgs production rate.}
     \label{fig:lumVsBeta}
 \end{figure}

Scans of the XFEL FWHM spot size,  r.m.s. electron longitudinal bunch density and electron final focus $\beta$ value are shown in Figs.~\ref{fig:lumVsAgamma},~\ref{fig:lumVsSigez}, and \ref{fig:lumVsBeta}, respectively.  In addition to illustrating the sensitivity of the luminosity to laser waist, electron bunch length and electron spot size, the plots demonstrate the impact of different Compton processes on the luminosity.  Note the large effect from the Bethe-Heitler process $e^- \gamma_0 \rightarrow e^- e^+ e^-$ in Figs.~\ref{fig:lumVsSigez} and \ref{fig:lumVsBeta}. 

Figs.~\ref{fig:lumVsAgamma} illustrates an interesting interplay between different effects as the photon density is increased.  As the photon density is increased
(i.e., ${\rm a}_{\gamma FWHM}$ is made smaller) the Compton conversion efficiency increases but the rate for
Bethe-Heitler production $e^-\gamma_0\rightarrow e^-e^+e^-$ and the non-linear
QED parameter $\xi^2$ also increase.  The result is a flat distribution in the Higgs production rate versus ${\rm a}_{\gamma FWHM}$ when all effects are included (blue curve).

 \subsubsection{Beam Extraction}
 
\begin{figure}
\centering
     \includegraphics[width=0.85\textwidth]{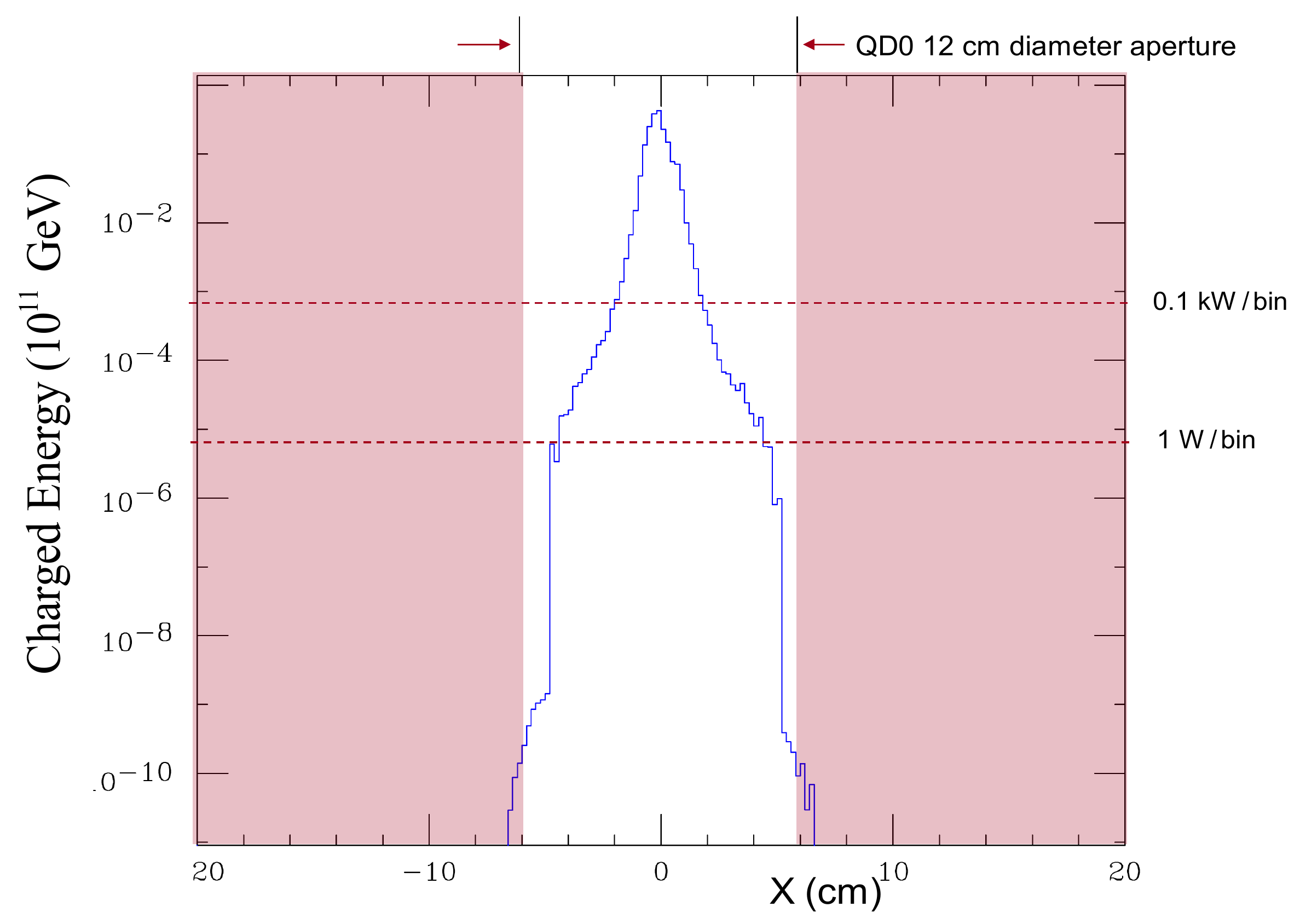}
     \caption{ Charged energy per beam per bunch crossing versus position transverse to the beam (X) at the distance $L^*=1.5$~m downstream of the IP assuming a 5~Tesla solenoid field.}
     \label{fig:evsx150}
 \end{figure}

\begin{figure}
\centering
     \includegraphics[width=0.85\textwidth]{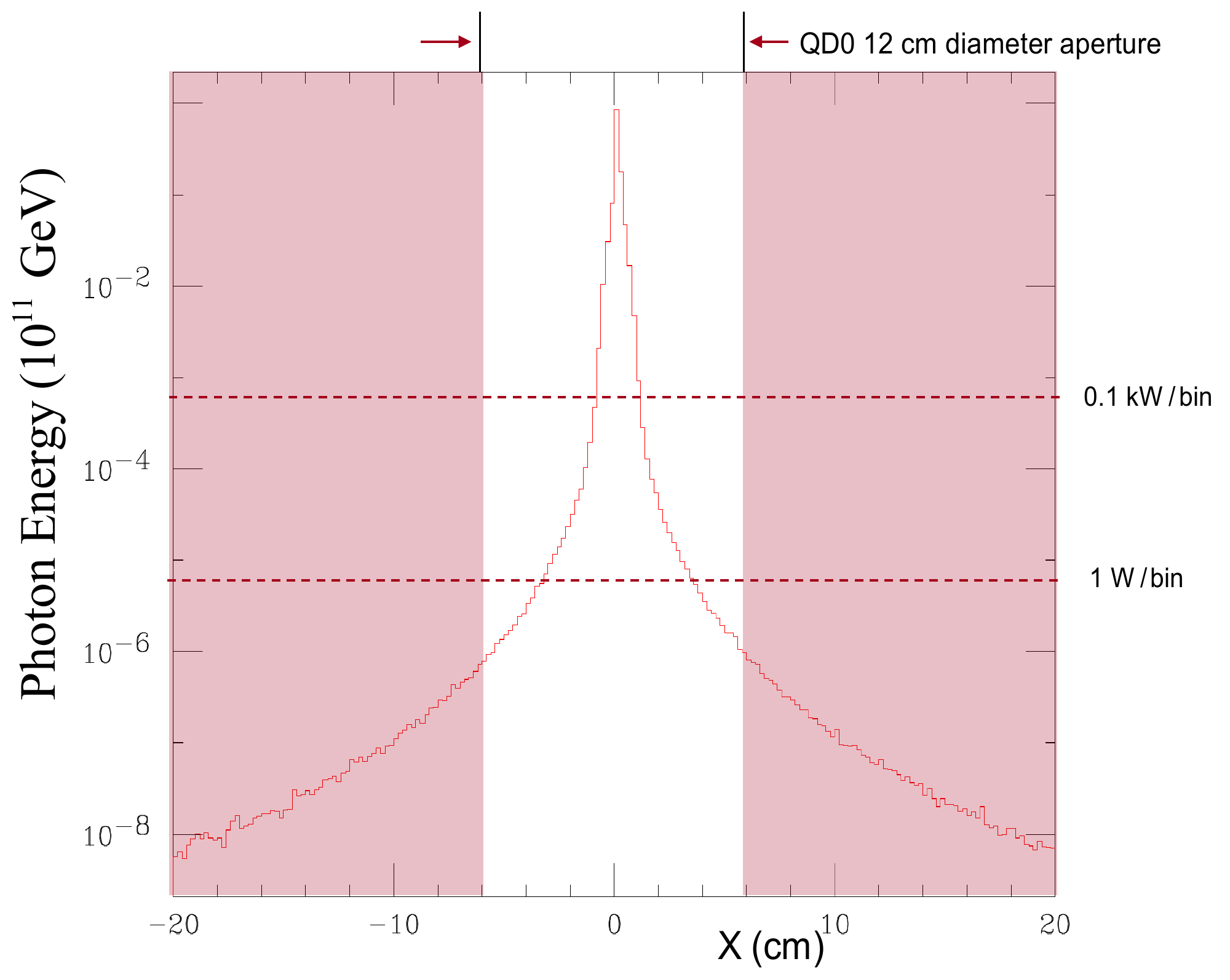}
     \caption{ Neutral energy per beam per bunch crossing versus position transverse to the beam (X) at the distance $L^*=1.5$~m downstream of the IP.}
     \label{fig:ephvsx150}
 \end{figure}

 The charged and neutral energy profiles several meters downstream of the IP, along with the strong anticorrelation between luminosity and crossing angles at both the IP and IPC, favor a nearly head-on collision.  A small crossing angle of 2~mrad was considered
 at one point for the ILC~\cite{Nosochkov:2005ip}, and this will serve as the starting point for the XCC beam extraction design.  The QD0 aperture in~\cite{Nosochkov:2005ip} was 9~cm in diameter.  The XCC assumes a 12~cm diameter aperture.  Figs.~\ref{fig:evsx150} and
 \ref{fig:ephvsx150} show the energy deposition per beam per bunch crossing as a function of position transverse to the beam at the distance $L^*=1.5$~m downstream of the IP assuming a 5~Tesla solenoid field.  In stable operation the energy deposition on QD0 is a few Watts, assuming no masking between the IP and QD0.  The energy of the photons striking QD0 in the region $|X|>6.0$~cm is about 1~MeV. 
 
\section{Energy upgrade to \texorpdfstring{$\sqrt{s}=380$}{Lg}~GeV}
\subsection{Introduction}

\begin{figure}
\centering
     \includegraphics[width=0.75\textwidth]{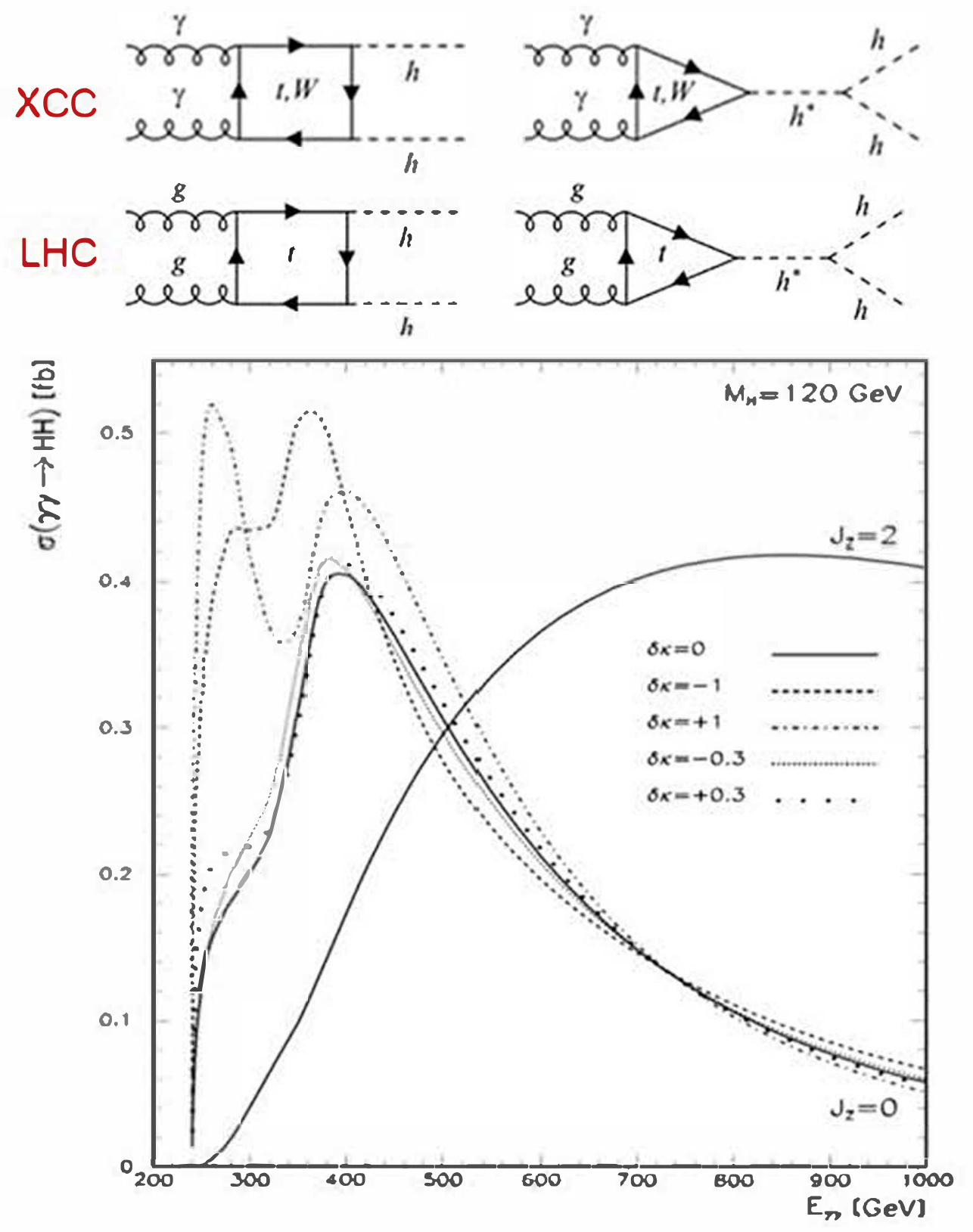}
     \caption{$\sigma(\gamma\gamma\rightarrow HH)$ versus $\gamma\gamma$ center-of-mass energy for various anomalous Higgs self coupling values $\delta\kappa$ (from reference\cite{Belusevic:2004pz}). }
     \label{fig:hhinter}
 \end{figure}

The XCC is ideally suited to study the Higgs self-coupling using 
$\gamma\gamma\rightarrow HH$ at $\sqrt{s}=380$~GeV. At 0.4~fb the cross-section\cite{Belusevic:2004pz} for $\gamma\gamma\rightarrow HH$ at $\sqrt{s}=380$~GeV is twice that of $e^+e^-\rightarrow ZHH$ at $\sqrt{s}=500$~GeV.
In addition, the $HH$ final state is simpler than the
$ZHH$ final state, which could carry some additional advantages.  N.B.,the
associated $Z$ boson in $e^+e^-$ production of the Higgs is great for measurements such as $\Gamma_{ZZ}$ and $\Gamma_\textrm{invisible}$, but can be a complication in other instances. Finally, the cross section $\sigma(\gamma\gamma\rightarrow HH)$ exhibits an interesting dependence on the  anomalous Higgs self-coupling $\delta\kappa$, as shown in Figure~\ref{fig:hhinter}. For example,  the cross section difference between $\delta\kappa=+1$ and $\delta\kappa=0$ is much larger for $\sigma(\gamma\gamma\rightarrow HH)$ than for $\sigma(gg\rightarrow HH)$\cite{Baglio:2012np}, due to the presence of both the $W$~boson and top quark in the Feynman diagram loops for $\gamma\gamma\rightarrow HH$.


\subsection{Energy upgrade parameters}
The parameters for the energy upgrade to $\sqrt{s}=380$~GeV are shown in table~\ref{tab:design380}. The final focus $\beta^*$ is reduced by a factor of three and the emittance is reduced by a factor of two. The X-ray energy is doubled, and the pulse length is cut in half to 10~$\mu$m.  The Compton collision $x$ parameter is maintained at $x=1000$, which means that the XFEL photon energy has been reduced from 1~keV to 0.34~keV; the 0.34 keV photon energy will clearly lead to X-ray focusing challenges, offset somewhat by the factor of three larger waist of 214~nm FWHM.   There is not yet a design for the 0.34~keV  XFEL and therefore some of the XFEL entries in table~\ref{tab:design380} are blank.

\begin{table}[htbp]
\begin{center}
\caption{\label{tab:design380} Summary of design parameters for $\sqrt{s}=380$~GeV.}
\begin{tabular}{ |l|l|||l|l|c| }
\hline
Final Focus parameters  & Approx.~value & XFEL parameters & Approx.~value \\
\hline
\hline
Electron energy & 190 GeV & Electron energy & 31 GeV \\ 
Electron beam power & 2.13 MW & Electron beam power &  0.34 MW \\
$\beta_x/\beta_y$ & 0.010/0.010 mm & Normalized emittance & 60 nm \\ 
$\gamma\epsilon_x/\gamma\epsilon_y$ & 60/60 nm & RMS energy spread $\langle\Delta\gamma/\gamma\rangle$ &  0.05\% \\ 
$\sigma_x/\sigma_y$ at $e^-e^-$ IP & 1.3/1.3 nm & Bunch charge & 1 nC \\ 
$\sigma_z$ & 10 $\mu\rm{m}$  & Linac-to-XFEL curvature radius & 133 km  \\ 
Bunch charge & 1 nC &  Undulator B field &  -  \\ 
Bunches/train at IP & 93 & Undulator period $\lambda_u$ &  - \\ 
Train Rep. Rate at IP & 120 Hz & Average $\beta$ function &  - \\
Bunch spacing at IP & 5.2 ns &  x-ray $\lambda$ (energy) &  3.6 nm (0.34 keV) \\
$\sigma_x/\sigma_y$ at IPC & 5.2/5.2 nm & x-ray pulse energy &  1.4 J  \\
$\mathcal{L}_\textrm{geometric}$ & $1.8\times 10^{36}\ \textrm{cm}^2\ \textrm{s}^{-1}$ & rms pulse length &  10 $\mu\rm{m}$  \\
$\delta_E/E$ & 0.1\% & $a_{\gamma x}$/$a_{\gamma y}$ (x/y waist) & 64.4/64.4~nm\\
$L^*$ (QD0 exit to $e^-$ IP) &  1.5m  & non-linear QED $\xi^2$ & 0.40  \\ 
$d_{cp}$ (IPC to IP) & $40\ \mu$m  & & \\
QD0 aperture & 12 cm diameter & & \\
Accel. gradient & 120 MV/m & & \\  \midrule
Site parameters & Approx. value & & \\ \midrule
crossing angle & 2 mrad & & \\
total site power & 140 MW & & \\
total length & 4.2 km & & \\
\hline
\end{tabular}
\end{center}
\end{table}

\subsection{Physics with \texorpdfstring{$\gamma\gamma\rightarrow HH$ and $\gamma\gamma\rightarrow t\bar{t}$}{Lg}}

Twenty year running scenarios for ILC/C$^3$ and XCC are shown in figure~\ref{tab:higgsfactories}. In Stage~II, ILC/C$^3$ and XCC produce $2\times 10^6$ and $2.9\times 10^6$ $t\bar{t}$ events, respectively.  Of greater interest is the fact that XCC produces more than twice as many double Higgs events for the same beam power as C$^3$, and for considerably less beam power than ILC.

 \begin{table}
    \begin{center}        
    \begin{tabular}{    l   | c c } \toprule
     & ILC/C$^3$ & XCC \\
     Colliding Particles   & $e^+e^-$ & $\gamma\gamma$ \\ \midrule
    Stage I:  & & \\ 
     \quad $\sqrt{s}$ (GeV) & 250 &  125 \\
     \quad Luminosity (fb$^{-1}$) & 2000 & 460 \\
     \quad Beam Power (MW) & 5.3 /\ 4.0 & 4.0 \\
     \quad Run Time (yr) & 10 & 10 \\ 
   \# Single Higgs  & $0.5\times 10^6$ & $1.3\times 10^6$ \\          \midrule
        Stage II: & & \\ 
     \quad $\sqrt{s}$ (GeV) & 550 &  380 \\
     \quad Luminosity (fb$^{-1}$) & 4000 & 4900 \\
     \quad Beam Power (MW) & 11 /\ 4.9 & 4.9 \\
     \quad Run Time (yr) & 10 & 10 \\ \midrule
    \# Single Higgs (I+II)  & $1.5\times 10^6$ & $1.3\times 10^6$ \\
    \# Double Higgs  & 840 & 1800 \\
    \# $t\bar{t}$  & $2.0\times 10^6$ & $2.9\times 10^6$ \\
   \bottomrule
    \end{tabular}
\end{center}
        \caption{ \label{tab:higgsfactories} 
        Running scenarios for ILC/C$^3$ and XCC including $\sqrt{s}$, luminosity, and beam power.  The number single Higgs, double Higgs and $t\bar{t}$ events are also indicated. }
 \end{table}

\begin{figure}
\centering
     \includegraphics[width=0.75\textwidth]{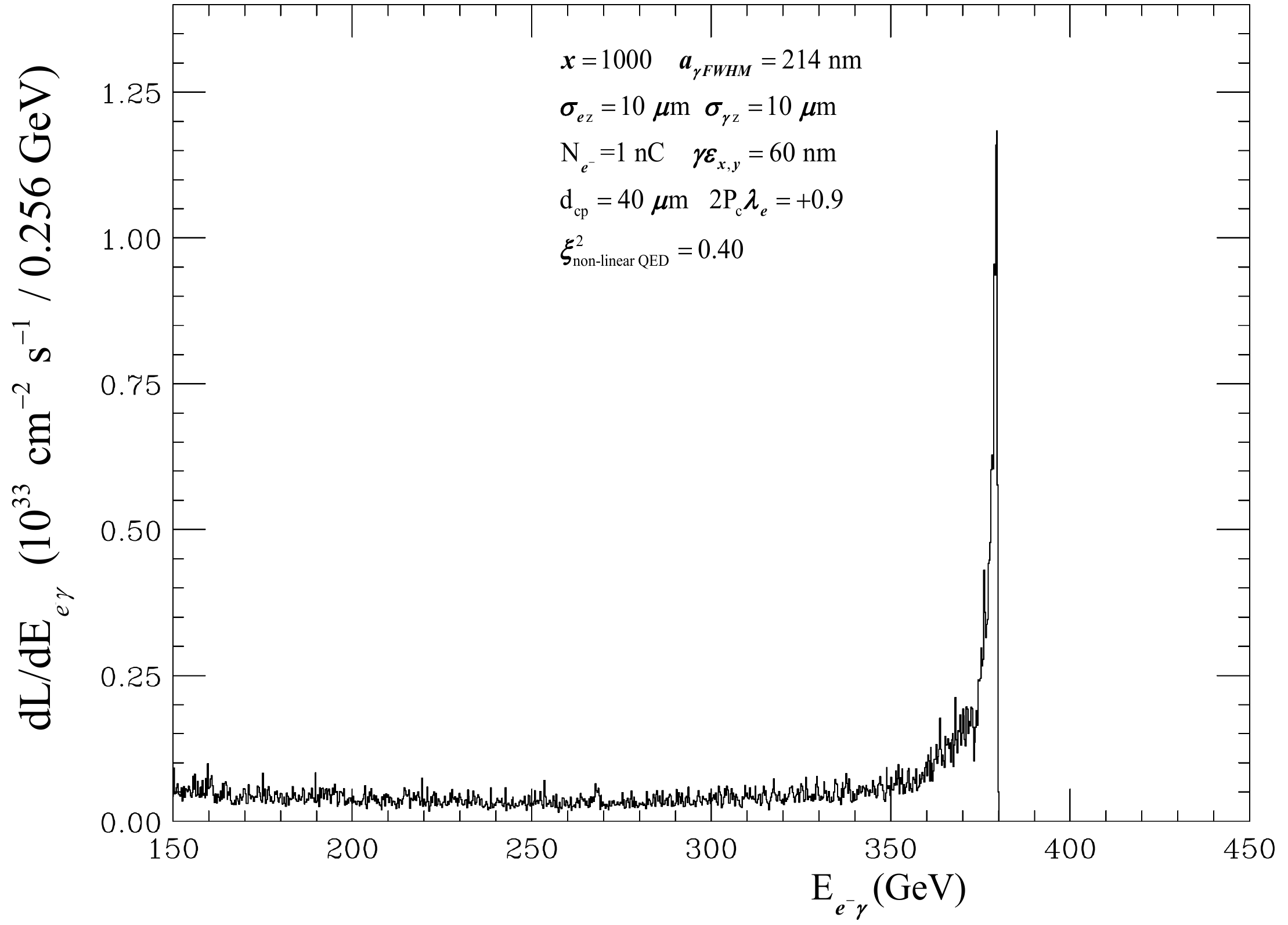}
     \caption{$e^-\gamma$ luminosity as a function of $e^-\gamma$ center-of-mass energy $E_{e^-\gamma}$ for x=1000 \& $2P_c\lambda_e=+0.9$ as calculated by the  CAIN MC, where $P_c$ and $\lambda_e$ are the helicities of the laser photon and electron, respectively. This luminosity is a byproduct of the primary $\gamma\gamma$ collision program at $\sqrt{s}=380$~GeV. The x-ray waist radius $a_\gamma$ at the Compton interaction point and the electron (x-ray) beam r.m.s longitudinal sizes, $\sigma_{ez}$ ($\sigma_{\gamma z}$), are indicated. The 0.1\% electron beam energy spread, linear QED Bethe-Heitler scattering and non-linear QED effects in Compton and Breit-Wheeler scattering are included in the CAIN simulation.
     }
     \label{fig:egamspec}
 \end{figure}

Beyond the study of $\gamma\gamma\rightarrow HH$ and $\gamma\gamma\rightarrow t\bar{t}$, another important aspect of XCC physics at $\sqrt{s}=380$~GeV is 
the measurement of $e^-\gamma\rightarrow e^- H$.  As mentioned in Section~\ref{sec:higgsphysics} this reaction can be used to measure the absolute partial width $\Gamma_{\gamma\gamma}$ which in turn provides model independent Higgs coupling measurements at XCC.
The $e^-\gamma$ luminosity spectrum, acquired as a byproduct  from running $\gamma\gamma$ collisions at $\sqrt{s}=380$, is shown in figure~\ref{fig:egamspec}. The  $e^-\gamma\rightarrow e^- H$ signal is a monochromatic 169.4~GeV electron, predominantly in the forward direction.  The cross section for Higgs production in $e^-\gamma$ collisions at $\sqrt{s}=380$~GeV is $\sigma(e^-\gamma\rightarrow e^- H)=4.0$~fb, assuming forward detector coverage down to an angle of 3~mrad.  The luminosity in the top 1\% of the spectrum in figure~\ref{fig:egamspec} is $9.8\times 10^{33}  \rm{cm}^2\ \rm{s}^{-1}$ so that 13,000 
$e^-\gamma\rightarrow e^- H$ events with a final state electron energy within 1\% of the 169.4~GeV peak would be collected in a ten year period assuming $1.6\times 10^7$~seconds per year.
A comparison of Higgs physics results at ILC/C$^3$ and XCC for 20 years of running is shown in table~\ref{tab:higgscouplings}.

 \begin{table}

    \begin{center}
    \begin{tabular}{    l  | c c } \toprule
    &  ILC/C$^3$ & XCC \\
     coupling $a$   &  $\Delta a$ (\%) &   $\Delta a$ (\%)  \\ \midrule
      $HZZ$ & 0.38 & 0.94 \\
      $HWW$ & 0.37 & 0.94  \\ 
      $Hbb$ &  0.60 & 0.95  \\ 
      $H\tau\tau$ & 0.77 & 0.99  \\ 
      $Hgg$ & 0.96 & 1.2 \\ 
      $Hcc$ & 1.2 & 1.2 \\ 
      $H\gamma\gamma$ & 1.0 & 0.44  \\ 
      $H\gamma Z$ & 4.0 & 1.5 \\ 
       $H\mu\mu$ & 3.8 & 3.5 \\
      $Htt$ & 2.8 & 4.6 \\
      $HHH$ & 20 & 14$^*$ \\  \midrule
     
      $\Gamma_{\rm tot}$ & 1.6 & 2.4 \\
      ${\Gamma_{\rm inv}}^\dagger$ & 0.32 & -- \\ 
      ${\Gamma_{\rm other}}^\dagger$ & 1.3 & 1.5 \\ 
              \bottomrule

    \multicolumn{2}{l}{\footnotesize{$^\dagger$~95\% C.L. limit}} \\
   \multicolumn{2}{l}{\footnotesize{$^*$assumes XCC error 
   is ILC/C$^3$ value scaled by $1/\sqrt{N_{HH}}$ }}\\

    \end{tabular}
        
        \end{center}

        \caption{ \label{tab:higgscouplings} Model independent Higgs coupling precision for the full 20 year programs at ILC/C$^3$ and XCC, as calculated with an Effective Field Theory (EFT) Higgs coupling fitting program\cite{Barklow:2017suo,Barklow:2017awn}. The errors on 
$\sigma(\gamma\gamma\rightarrow H)\times \rm{BR}(H\rightarrow X)$ are taken from~\cite{Bambade:2019fyw}, and are assumed to be the same for ILC and XCC for all decay modes except $H\rightarrow \gamma\gamma$
and $H\rightarrow\gamma Z$, which are characterized by monochromatic photons of energy 62.5~GeV and 29.3~GeV, respectively, at XCC.}

 \end{table}

\section{Design Challenges}
\subsection{Electron Final Focus}
The  final focus design differs from than that of ILC \& CLIC in 2 key aspects, each of which raises concerns which need further studies to address:
\begin{itemize}
    \item 	Round beams at the IP leads to the preference of a final triplet instead of final doublet configuration. The required angular dispersion at the IP is about double that required for ILC/CLIC to achieve the same dispersion at the sextupole locations. This will have an adverse effect on the momentum acceptance of the extraction line and may lead to increased detector backgrounds.
	\item	The requested IP beta functions are 0.03 mm in both planes. This should be compared with 11 x 0.48 mm for the baseline ILC design (and a corresponding design tested at the ATF2 facility). The much smaller $\beta^*$ values here generate significantly higher chromatic distortions, requiring stronger sextupole corrections. This in turn requires more finely tuned 3rd, 4th+ order corrections to compensate for the sextupoles. Experience from CLIC tuning studies and operational experience at ATF2 has shown that tolerances become rapidly tighter (magnet field quality and positional tolerances) as $\beta^*$  is lowered below ILC values, and online tuning becomes harder and takes longer. Also, operational experience at ATF2 showed that tuning becomes more difficult with smaller $\beta^*_x$  : $\beta^*_y$ ratios (where the smallest spot sizes were only accomplished at 10X design $\beta^*_x$). With this in mind, a careful study of the tolerances of this, modified, final focus design is important to understand the ramifications on expected delivered luminosity.
	\end{itemize}
	
 \subsection{X-ray optics}
 The basic strategy for the X-ray optics design was outlined in section~\ref{xrayoptics}.   No clear solution for 0.7~J pulse energy
 and 1~keV photon energy was identified in this study.  Further R\&D is required to determine if there are fundamental limits to the focusing of 1~keV photons with  $\sim 1$~J pulse energies, or if, for example,
 the development of even longer FEL quality KB substrates ($> 2$ meters) can solve the problem.  
 
\subsection{Timing and positional stability}
R\&D is required to determine the timing and positional stability specifications for the electron and laser beams, and to deliver technology that meets these specifications.

\subsection{Longitudinal wake fields and electron energy spread}
Initial calculations indicate that the short 20$\mu$m rms pulse length and  the small aperture of the C-band structures combine to produce large longitudinal wake field effects that create an energy spread beyond the tolerance of the final focus system.  Work is ongoing to identify solutions  without abandoning C-band structures.

\section{High Brightness LCLS-NC Demonstrator Project}
The required key technologies for the XCC are:
\begin{enumerate}
    \item 1 nC high brightness cryogenic RF gun
    \item 700 mJ/pulse XFEL
    \item 1 keV X-ray optics with 700 mJ/pulse
    \item C$^3$ Acceleration of GeV-class electron beam
\end{enumerate}

Key XCC technology (4) is discussed in the
C$^3$ demonstration document~\cite{Nanni:2022oha}.

Key XCC technologies (1), (2), and (3) can be tested by building a 1 nC high brightness cryogenic RF injector for LCLS-NC~\cite{frisch:2021}.  The construction of such an injector would help demonstrate
key technology (1).  When such an injector is incorporated into LCLS-NC, the upgraded x-ray laser could be used to demonstrate key technologies~(2) and (3).  Furthermore, such a high brightness upgrade to LCLS-NC could open up exciting new research opportunities in photon science.

\subsection{1nC 120 nm-rad cryogenic RF electron gun}
A cryogenic copper RF gun operating at C-band with 0.1 nC/pulse and 45 nm emittance is being developed for an ultra-compact XFEL\cite{Robles:2021ixk}. Scaling arguments indicate that an injector with 1nC/pulse and 120~nm emittance should be possible.  A design study performed several years ago for a cryogenic copper RF gun operating at S-band (TOPGUN) demonstrated that an emittance of 200~nm-rad could be achieved with 1~nC/pulse~\cite{Cahill:2017kwh}\cite{Cahill:2017thesis}.  The development of this gun is a prerequisite for the LCLS-NC high brightness upgrade discussed in the following section. Additional details on high bunch charge cryogenic copper RF gun development can be found in the
C$^3$ demonstration document~\cite{Nanni:2022oha}.

\subsection{LCLS-NC performance}
\begin{figure}
\centering
     \includegraphics[width=0.65\textwidth]{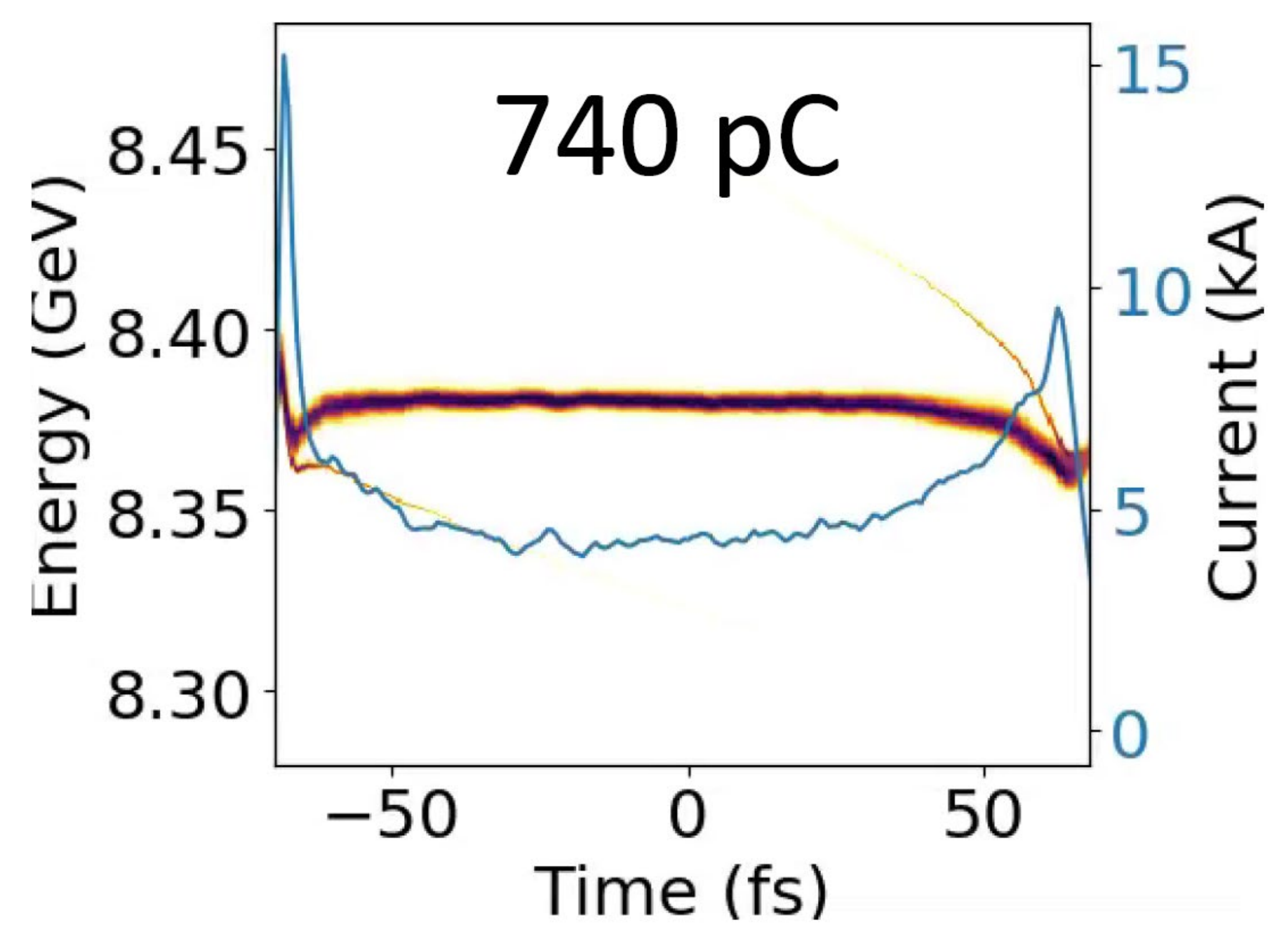}
     \caption{Electron energy and current profile at Linac exit in the high brightness upgrade of LCLS-NC.}
     \label{fig:energycurrentexit}
 \end{figure}

The performance of LCLS-NC assuming an injector with 1~nC charge per pulse,  120 nm-rad emittance and $40\ \mu$m bunch length has been simulated using  ELEGANT\cite{osti_761286} for the Linac and GENESIS for the soft x-ray undulator.  The beam after the injector was created by taking a typical 250~pC beam, scaling the projected emittance from 0.49 to 0.12~$\mu$m in each plane, and increasing the charge per simulation macro particle.  The 1 nC bunch was collimated to 0.74 nC in the first bunch compressor to cut horns.  Linac phases were adjusted to accelerate the beam to 4.5 GeV before the second bunch compressor and compress the beam to 4.5~kA. The compressed beam was accelerated on crest in the first 61 cavities of the third linac to achieve a energy of 8.4~GeV, and the final 117 cavities were set to -80.6 degrees to remove a 15 MeV chirp across the beam. The electron energy and current profile at the exit of the Linac is shown in Fig.~\ref{fig:energycurrentexit}.

Seeded FEL simulations assume a monochrome 50~kW, 1~keV seed approximating a seed from self-seeding to achieve a high brightness x-ray source.  Upon entering the soft x-ray undulator line, the electron beam energy is 8.4~GeV, which is the highest energy beam that can lase at 1~keV in this undulator line (constant period of 3.9~cm and max undulator normalized vector potential of 5.7).  The beam's horizontal and vertical projected emittances of 1.9~$\mu$m and 0.34~$\mu$m, respectively, are dominated by contributions from the horns at the head and tail of the beam, yet the projected core emittance (estimated from the middle 20~fs of beam) remains 0.11~$\mu$m in both planes.  With the standard undulator FODO lattice, the electron beam transverse rms of $11\ \mu$m implies an x-ray waist of $22\ \mu$m, and Rayleigh length of 1~m.  This implies significant diffraction within a 0.8~m gain length (estimated via \cite{Xie2000}), and a shot noise power of 7~kW which is 14\% of the desired seed power of 50~kW which may lead to significant SASE breakthrough. Reducing the FODO quad gradients to 21\% of normal increases the electron beam horizontal rms width to $42\ \mu$m, implying an x-ray Rayleigh range of 5~m, which is significantly longer than the 1.2~m gain length. This also reduces the shot noise to 2.5 kW, or 5\% of the seed power. 

\begin{figure}
\centering
     \includegraphics[width=0.65\textwidth]{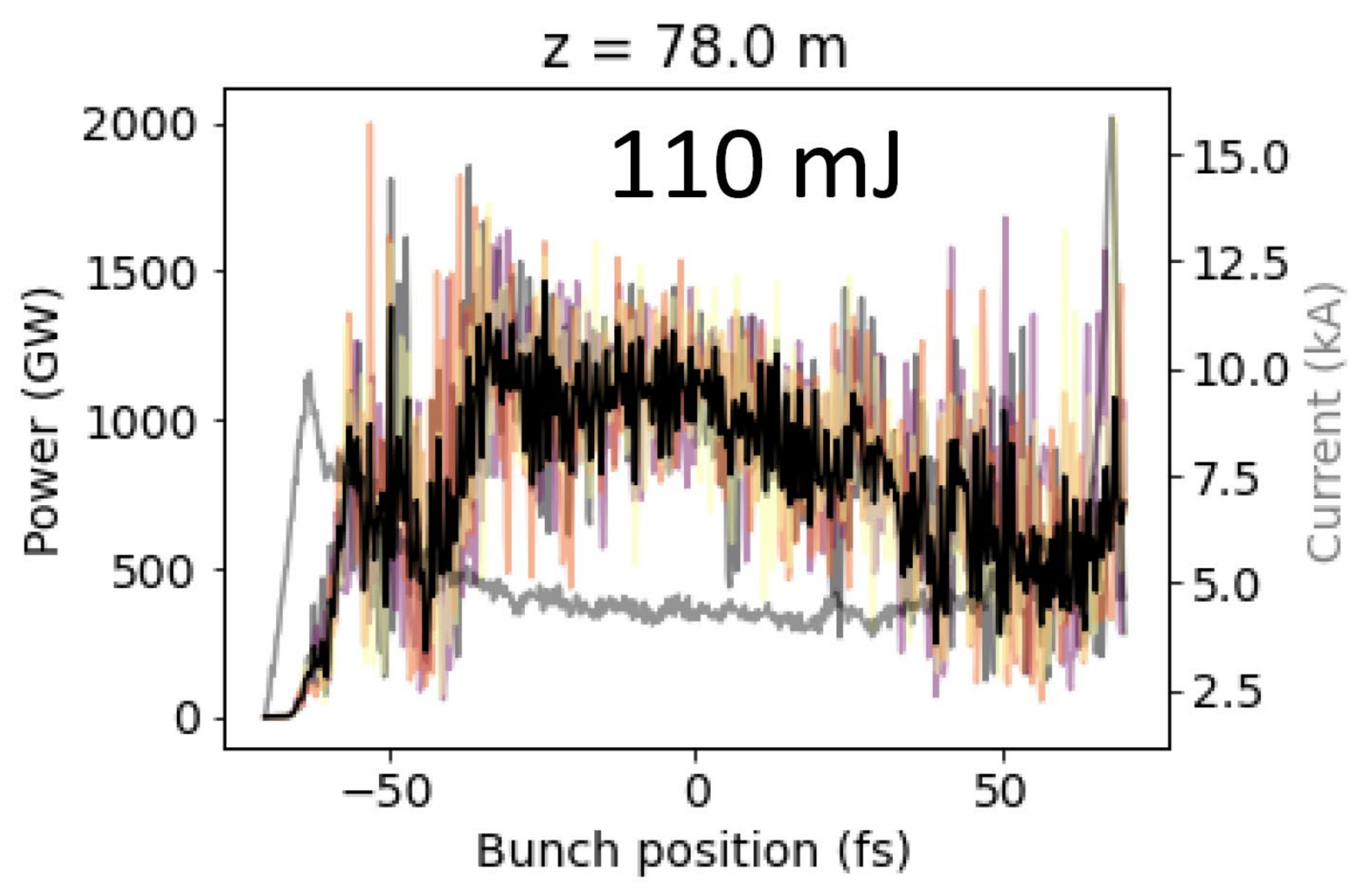}
     \caption{X-ray power profile in the high brightness upgrade of LCLS-NC assuming 21 undulators downstream of the SXRSS chicane.}
     \label{fig:xraypower}
 \end{figure}

\begin{figure}
\centering
     \includegraphics[width=0.65\textwidth]{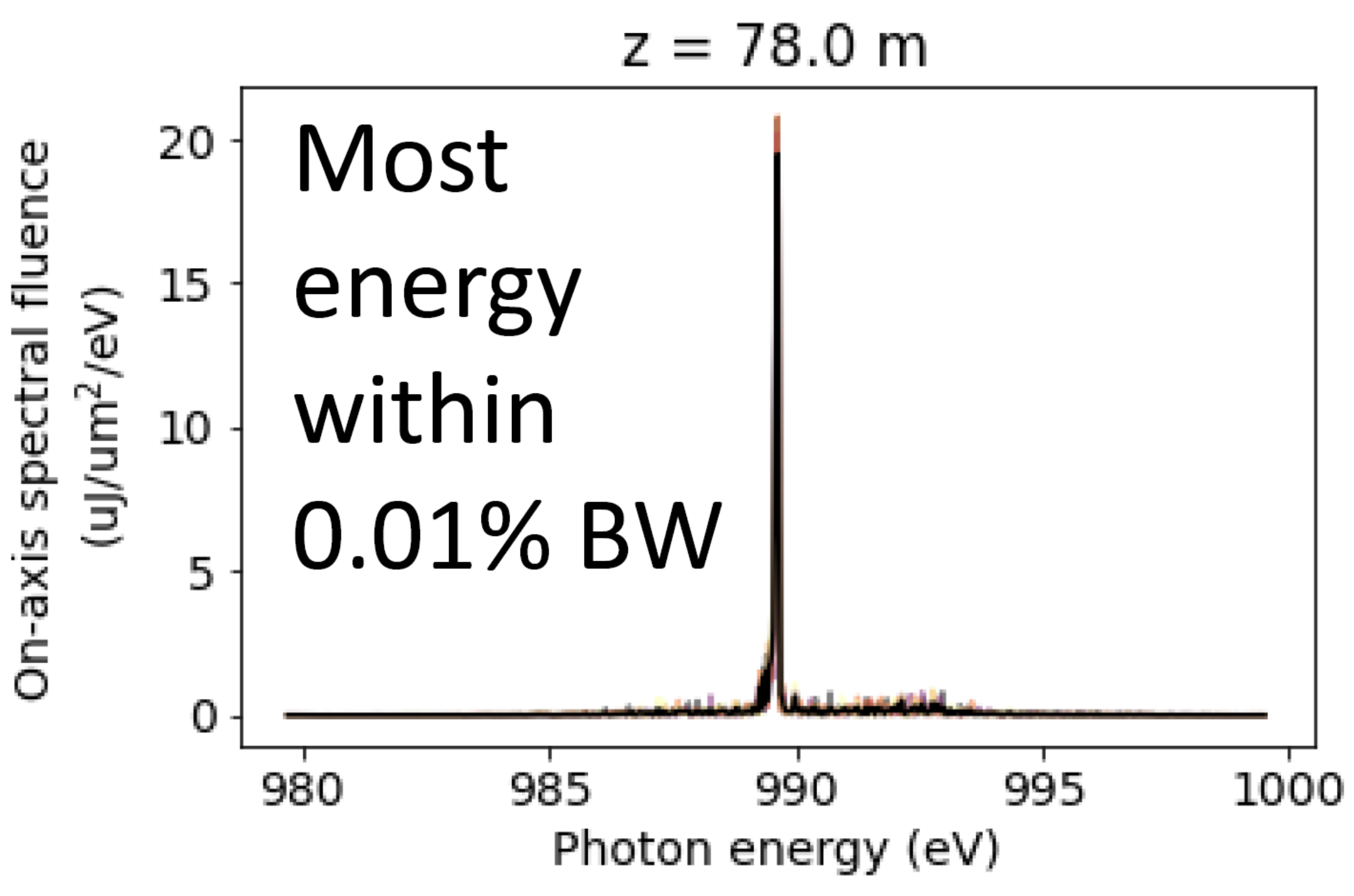}
     \caption{X-ray spectral fluence in the high brightness upgrade of LCLS-NC assuming 21 undulators downstream of the SXRSS chicane.}
     \label{fig:spectralfluence}
 \end{figure}

The seed power for Soft X-ray Self Seeding (SXRSS) is limited to 50~kW to avoid damage to the spectral collimating optics in the SXRSS monochromator. The seeded FEL saturates in about 6 undulators (23.4~m), after which the remaining undulators are quadratically tapered to reduce the undulator field strength by 5.6\% to enhance the peak power to over 1~TW.  After a total of 82~m (21 undulators following the SXRSS chicane) the energy per pulse is 110~mJ with the power profile shown in Fig.~\ref{fig:xraypower}. With the current 12 undulators following the SXRSS chicane, the energy per pulse is 33~mJ. The spectral fluence is shown in Fig.~\ref{fig:spectralfluence} where the bandwidth is less than 0.01\%~FWHM.

The 0.01\% bandwidth is much smaller than that required for XCC optical studies, but is important for photon science applications.   Most photon science applications require a much shorter pulse of 10~fs or less.  This can be accomplished with a slotted foil in a dispersive region, with a linear loss in pulse energy versus pulse length.  Furthermore, enhanced SASE could also benefit from low emittance, high current beams to enhance the power of sub-femtosecond soft X-ray pulses.



\clearpage

\acknowledgments

This work was supported in part by the U.S. Department of Energy (DOE) (Contract No. DE-AC02-76SF00515).  We would like to acknowledge the contributions of Joe Duris, Joe Frisch, and David Fritz to this research.




\bibliographystyle{JHEP}
\bibliography{biblio.bib}






\end{document}